\documentclass{aa}  
\usepackage{graphicx}
\usepackage{txfonts}
\usepackage{subfigure}
\usepackage[figuresright]{rotating}
\usepackage{textcomp}
\usepackage{txfonts}
\def\degr{\hbox{$^\circ$}}

\def\aap{A\&A}

\def\apjs{ApJS}

\usepackage{natbib} 
\bibpunct{(}{)}{;}{a}{}{,}

\begin{document}

\title{The Struve-Sahade effect in the optical spectra of O-type binaries\thanks{Based on observations made at the European Southern Observatory (La Silla, Chile) and at the Observatoire de Haute Provence (France).}}
\subtitle{I. Main-sequence systems}
\author{N. Linder\inst{1} \and G. Rauw\inst{1}\fnmsep\thanks{Research Associate FNRS, Belgium} \and H. Sana \inst{2}\and M. De Becker\inst{1}\fnmsep\thanks{Postdoctoral Researcher FNRS, Belgium} \and E. Gosset\inst{1}\fnmsep$^{\star\star}$}

\offprints{N. Linder}

\institute{Institut d'Astrophysique et de G\'eophysique, Universit\'e de Li\`ege, B\^at. B5c, All\'ee du 6 Ao\^ut 17, B-4000 Li\`ege, Belgium\\
           \email{linder@astro.ulg.ac.be}
  \and
           European Southern Observatory, Alonso de Cordova 3107, Vitacura, Santiago 19, Chile\\
          }
\date{}

\abstract{}{We present a spectroscopic analysis of four massive binary systems that are known or are good candidates to display the Struve-Sahade effect (defined as the apparent strengthening of the secondary spectrum of the binary when the star is approaching, and the corresponding weakening of the lines when it is receding).} 
{We use high resolution optical spectra to determine new orbital solutions and spectral types of HD~165\,052, HD~100\,213, HD~159\,176 and DH~Cep. As good knowledge of the fundamental parameters of the considered systems is necessary to examine the Struve-Sahade effect. We then study equivalent width variations in the lines of both components of these binaries during their orbital cycle.}
{In the case of these four systems, variations appear in the equivalent widths of some lines during the orbital cycle, but the definition given above can any longer be valid, since it is now clear that the effect modifies the primary spectrum as much as the secondary spectrum. Furthermore, the lines affected, and the way in which they are affected, depend on the considered system. For at least two of them (HD~100\,213 and HD~159\,176) these variations probably reflect the ellipsoidal variable nature of the system.}{}

\keywords{stars: individual: HD\,100213, HD\,159176, HD\,165052, DH\,Cep -- binaries: spectroscopic -- stars: fundamental parameters}

\maketitle

\section{Introduction}

Massive early-type stars of spectral type O, because of their high temperature ($T_{\mathrm{eff}}$~$\ge$~$30000$~K) and luminosity ($\sim 10^5$~-~$10^6$~L$_\odot$), play a key role in the ecology of galaxies. Indeed, these objects are responsible for the ionisation of gaseous nebulae, and lose a huge quantity of matter in the form of stellar winds ($\sim 10^{-7}$-$10^{-5}$~M$_\odot$~yr$^{-1}$). A large fraction of these stars are actually binary or multiple systems. Whilst this fact allows us to determine some fundamental properties of these stars, it also introduces other unexpected effects. One of the possible signatures of the interactions between the system's components is the so-called {\itshape Struve-Sahade effect} (S-S effect, hereafter). This effect is defined as the apparent strengthening of the secondary spectrum of the binary when it is approaching, and the corresponding weakening of its lines when it is receding \citep{BGR99}. First reported by \citet{Struve37}, this effect casts some doubt on the accuracy of massive binary parameters, such as masses and luminosity ratios. The first tentative explanation, essentially qualitative, came from \citet{Struve50}, who associated this effect with gaseous streams sweeping the trailing side of the secondary and then obscuring the secondary star when the latter is receding (see also \citealt{Sahade59}). More recently, \citet{GBP97} proposed a completely different scenario. They argued that the bow shock produced by the encounter of the two stellar winds is deflected by the Coriolis force to a position above the leading atmosphere of the secondary star. Assuming that the wind of the secondary star is the weaker one, the secondary's surface would be heated by the proximity to the shock region and this would increase its contribution to the composite spectrum when it is approaching (deeper lines). The studies of \citet{Stickland97} and \citet{BGR99} on IUE data of three massive binaries (\object{HD~47\,129}, \object{AO~Cas} and \object{29~CMa}) did not allow them to determine the cause of the effect, but they concluded that these stars tell `a different tale', and that the Struve-Sahade effect may arise from several distinct mechanisms. More recently, the theoretical developments of \citet{Gayley02} and \citet{GTP07} have suggested that the Struve-Sahade effect could be linked to surface flows driven away from one star which would tend to compensate the rotational broadening of the component moving towards the observer (yielding narrower and deeper lines) and enhance the rotational broadening when the component is receding (broader and shallower lines). While preliminary results indicate that this new scenario can produce variations similar to those seen in the S-S effect, detailed comparison with observational data is still needed.

In this paper, we discuss a set of spectroscopic observations of four early-type binaries. Our sample consists of main-sequence O-type stars with orbital periods of a few days. We derive new orbital solutions for these systems and analyse the variability of the spectral lines considering the Struve-Sahade effect. 

\section{HD~165\,052}

\object{HD~165\,052} is a massive binary that probably belongs to the open cluster NGC~6530, in the Lagoon nebula \citep{VAJ72}. Its SB2 nature was first discovered by \citet{Conti74} and \citet{MC78} derived the first orbital solution from 20 optical spectra obtained with the Coud\'e-feed telescope at Kitt Peak. They found a slight eccentricity ($e$~=~0.06~$\pm$~0.04), a period of $6.14 \pm 0.002$ days and similar spectral types (O6.5~V) for both components. \citet{Morrison75} observed this system without detecting any photometric variation. Furthermore, \citet{SLK97} used IUE data to obtain a new orbital solution and the better quality of their data allowed them to find a period of 2.96 days. They assumed a zero eccentricity. Finally, \citet{AMB02} analysed 73 spectra of \object{HD~165\,052}, with different resolutions and spectral ranges (from 3900 to 7600 \AA). They confirmed the period found by \citet{SLK97} but found an eccentricity equal to $0.09 \pm 0.004$. They used their new values of $e$ and $P$ to recompute orbital solutions from older data and detected apsidal motion. In the X-ray domain, \citet{Corcoran96} detected a phase-locked variability in ROSAT archive data of \object{HD~165\,052}, but this was based on the wrong value of the orbital period. This fact was however confirmed by \citet{AMB02}, who reanalysed the data using their new period. They observed phase-locked variations of the X-ray flux with a rise of the emission near both quadratures and explained these modulations as a result of the presence of colliding winds in the system.

Observations of \object{HD~165\,052} were gathered at the European Southern Observatory (ESO), with the Fiber-fed Extended Range Optical Spectrograph (FEROS) mounted on the 1.5~m telescope at La Silla. Twenty one spectra were obtained between 1999 and 2002, in the [3750, 8700]~\AA~wavelength domain. Typical exposure times range from 7.5 to 20 minutes and the mean signal to noise ratio (SNR) of the spectra is 340. The spectral resolving power of FEROS is 48000. The journal of observations is presented in Table~\ref{tab:journal:165052}. The data were reduced using an improved version of the FEROS pipeline \citep[see][]{SHR03} working in the MIDAS environment. We mainly used the normalised automatically merged spectrum, and checked our results on the individual orders in case of doubts.

\begin{table}
\caption{Journal of the observations of \object{HD~165\,052}. The first column lists the heliocentric Julian date (HJD), the second provides the phase as computed from our orbital solution (see Table \ref{tab:orbit:165052}) and the third and fourth ones give the radial velocities used to compute the orbital solution respectively for the primary and secondary stars.}
\label{tab:journal:165052}
\centering
\begin{tabular}{c c r r}
\hline
\hline
HJD &$\Phi$ &$RV_1$~~~&$RV_2$~~~ \\    
$-$ 2 450 000 & &(km s$^{-1}$) & (km s$^{-1}$)\\
\hline
\multicolumn{4}{c}{}\\
1299.7250 &0.23 &88.3~~~ &  $-$100.8~~~\\
1300.7319 &0.57 &$-$56.5~~~ &    76.9~~~\\
1300.9264 &0.63 &$-$86.7~~~ &    96.6~~~\\ 
1301.9281 &0.97 &31.2~~~ &   $-$24.9~~~\\
1304.7434 &0.93 &$-$80.1~~~ &   103.7~~~\\
1304.7507 &0.93 &96.3~~~ &  $-$111.3~~~\\ 
1304.9309 &0.99 &$-$87.7~~~ &   106.8~~~\\
1323.8361 &0.39 & $-$93.7~~~ &   103.3~~~\\
1327.6014 &0.66 &7.4~~~ &    $-$1.7~~~\\  
1327.9127 &0.77 &15.0~~~ &   $-$12.5~~~\\ 
1670.7601 &0.78 &40.1~~~ &   $-$39.5~~~\\ 
1671.7225 &0.11 &17.9~~~ &   $-$25.8~~~\\
1672.7016 &0.44 &$-$15.9~~~ &    28.7~~~\\  
2335.8879 &0.86 &  $-$41.0~~~ &    56.9~~~\\
2336.8791 &0.19 &97.5~~~ &  $-$118.5~~~\\  
2337.8880 &0.53 &$-$54.9~~~ &    70.1~~~\\  
2338.8808 &0.87 &$-$46.1~~~ &    47.8 ~~~\\
2339.8848 &0.21 &96.9~~~ &  $-$111.2~~~\\  
2381.8324 &0.40 &6.6~~~ &    $-$5.7~~~\\
2382.8569 &0.75 &$-$79.7~~~ &   103.3~~~\\  
2383.8577 &0.09 &95.4~~~ &  $-$103.1~~~\\
\hline
\end{tabular}
\end{table}

\begin{figure}
  \resizebox{\hsize}{!}{\includegraphics{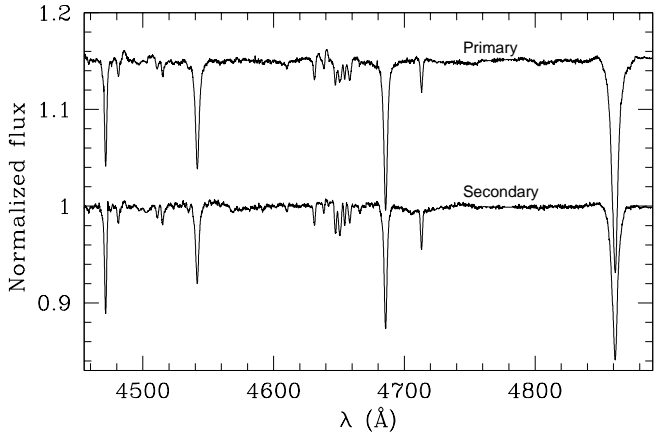}}
  \caption{Normalised disentangled spectra of \object{HD~165\,052}, between 4455 and 4895 \AA. The primary spectrum was vertically shifted by 0.15 units for clarity.}
  \label{fig:165052dis}
\end{figure}

\subsection{Orbital solution}

\begin{figure}
\centering
\resizebox{0.9\hsize}{!}{\includegraphics{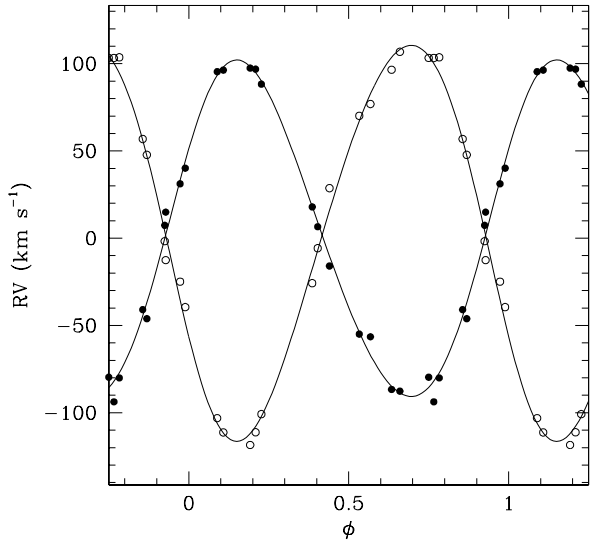}}
\caption{Radial velocity curve of \object{HD~165\,052} calculated from the orbital solution of Table~\ref{tab:orbit:165052}. Primary data are represented by filled symbols, while secondary data are represented by open symbols. The RVs have been determined through a disentangling of the primary and secondary spectra using the method of \citet{GL06}. $\Phi = 0$ corresponds to the phase of the periastron passage ($\Phi = 0.93$: secondary conjunction, $\Phi = 0.15$: primary's ascending node, $\Phi = 0.41$: primary conjunction, $\Phi = 0.70$: secondary ascending node).}
\label{fig:vrad:165052}
\end{figure}

In order to calculate a new orbital solution, we applied a disentangling method to separate the two components of the system while simultaneously determining their radial velocities \citep{GL06}. This consists of an iterative procedure that alternatively uses the spectrum of one component (properly shifted according to its radial velocity) to subtract it from the observed spectra and to calculate an approximation of the spectrum of the other component. This technique is more powerful than the commonly used method of fitting a Gaussian profile to the studied lines, and allows us to determine the radial velocities of both components at phases when the lines are heavily blended. The disentangled spectra of the components of \object{HD~165\,052} are shown in Fig.~\ref{fig:165052dis}. Due to some residual uncertainties on the normalisation of the observed spectra, some artefacts appear in the disentangled spectra as low frequency modulations of the continuum level. The values of the radial velocities (RVs) found with this method are listed in the third and fourth columns of Table~\ref{tab:journal:165052}. The lines used for this calculation are He I $\lambda$~4471, Mg II $\lambda$~4481, He II $\lambda$~4542, He II $\lambda$~4686, He I $\lambda$~4713 and H$_\beta$. The orbital solution itself was calculated with the program LOSP (Li\`ege Orbital Solution Package) which is based on the method of \citet{WHS67}, subsequently modified to be applied to SB2 systems \citep{RSG00,SGR06b}. Table~\ref{tab:orbit:165052} gives the main orbital elements computed for \object{HD~165\,052}, and the corresponding uncertainties. As a first guess we adopted the period from \citet{SLK97} and allowed it to vary during the computation of the solution. The results obtained are close to the parameters of \citet{AMB02}, except for a slightly larger $K_2$ and hence for somewhat larger masses. Fig.~\ref{fig:vrad:165052} shows the radial velocity curve of \object{HD~165\,052} as a function of the orbital phase.

Using the disentangled spectra, we determined the projected rotational velocities ($v\,\sin{i}$) of the primary and secondary star applying the Fourier method \citep[see][]{SDH07,Gray2005} to the profiles of the disentangled He~I~$\lambda$~4471, He~II~$\lambda$~4542 and H$_\beta$ line profiles. In this way, we determine values of $73 \pm 7$ and $80 \pm 7$~km~s$^{-1}$ for the primary and secondary component of \object{HD~165\,052} respectively. These values are similar, which suggests a synchronous rotation of the components since their spectral types, and hence radii, are very similar. Furthermore, when we introduced the parameters of the system in the formula of \citet{Tassoul1990}, we found a circularisation time of less than 33\,500 years. However, we kept the eccentric solution as the best orbital solution for \object{HD~165\,052} (such as \citealt{AMB02}), because of the better quality of the fit. Indeed, the rms of 9.0~km~s$^{-1}$ is improved to 6.2~km~s$^{-1}$ when we consider an eccentricity equal to~0.08 instead of~0. Finally, if we suppose that the angular velocity of the system is equal to the individual angular velocities of the stars (co-rotation), we find that the inclination of the orbit should be equal to 28\degr. This value is too large compared to the inclinations found by \citet{SLK97} and \citet{AMB02} and leads to surprisingly small absolute masses. There is thus probably no synchronous co-rotation in \object{HD~165\,052}.

\subsection{Spectral type determination \label{std165}}

A quantitative spectral classification for O-type stars has been developed by Conti \citep[][and subsequent articles]{CA71} and refined by \citet{Mathys88, Mathys89}. This criterion is based on the ratio between the equivalent widths (EW) of He~I~$\lambda$~4471 and He~II~$\lambda$~4542. The equivalent widths were directly measured on the spectra after a standard deblending of the lines, by fitting two Gaussian profiles to the data. This leads to an O6 spectral type for the primary and O6.5 for the secondary. The uncertainties on these numbers (at least partially due to the S-S effect, see Sect.~\ref{ss_165052}) correspond to a range for the primary between O6 and O6.5, and for the secondary between O6 and O7. As a second step, equivalent widths were measured on the disentangled spectra and the corresponding spectral types are O6.5~+~O7. In agreement with \citet{Mathys88}, the fact that the He II $\lambda$~4686 line is seen in absorption, without any hint of an emission component, suggests that the luminosity class of both stars is V. For spectral types O7 and earlier, \citet{Mathys88} proposed EW$_{\lambda}$(4686) = 0.56 \AA~as a demarcation between the giant and main-star luminosity classes. If we take the optical brightness ratio of the system into account ($\frac{L_1}{L_2} = 1.55$), our measurements of equivalent widths on the \object{HD~165\,052} FEROS spectra indicate values of 0.79 and 1.0 \AA~for the primary and the secondary, respectively, thus confirming the main-sequence luminosity class of the components. This optical brightness ratio has been determined by comparing the observed EW ratio between the primary and secondary star to the same ratio evaluated for a sample of stars with the same spectral type \citep{CA71,Conti73}. The comparison was performed for the He~II~$\lambda\lambda$~4200,4542 and 4686 lines, and $\frac{L_1}{L_2} = 1.55$ is thus valid in the optical domain. Finally, spectra in Fig.~\ref{fig:165052dis} show the N~III~$\lambda$$\lambda$~4634,4640,4641 lines in emission, wich leads to an O((f)) spectral type.

\begin{table}
\caption{Orbital solution of HD165052. $T_0$ refers to the time of periastron passage. $\gamma$, $K$ and $a\;sin i$ are respectively the systemic velocity, the amplitude of the radial velocity curve, and the projected separation between the centre of the star and the centre of mass of the binary system. Quoted uncertainties are the 1$\sigma$ error bars.}
\label{tab:orbit:165052}
\centering
\begin{tabular}{lrclcrcl}
\hline
\hline
Parameter           &\multicolumn{3}{c}{Primary}& &\multicolumn{3}{c}{Secondary}\\
\hline
\multicolumn{8}{c}{}\\
 $P$ (days)& &  &\multicolumn{3}{c}{2.95515 $\pm$ 0.00004}& & \\
 $e$ & & &\multicolumn{3}{c}{0.081 $\pm$ 0.015}& & \\
 $\omega$ (\degr)& & & \multicolumn{3}{c}{298.0 $\pm$ 10.2} & &\\
 $T_0$ (HJD)& & &  \multicolumn{3}{c}{2451299.053 $\pm$ 0.081}& & \\
 $\gamma$ (km s$^{-1}$)&   \multicolumn{3}{c}{2.1 $\pm$ 1.2}& &\multicolumn{3}{c}{1.4 $\pm$ 1.3} \\
 $K$ (km s$^{-1}$)&  \multicolumn{3}{c}{96.4 $\pm$ 1.6}& &\multicolumn{3}{c}{113.5 $\pm$ 1.9}\\
 $a$ $sin i$ ($R_{\odot}$)&  \multicolumn{3}{c}{5.6 $\pm$ 0.1}& &\multicolumn{3}{c}{6.6 $\pm$ 0.1} \\
 $M$ $sin^3 i$ ($M_{\odot}$) &    \multicolumn{3}{c}{1.5 $\pm$ 0.1}& &\multicolumn{3}{c}{1.3 $\pm$ 0.1}\\
 $Q$ ($  M1/M2 $) & & & \multicolumn{3}{c}{1.18 $\pm$ 0.02}& &\\
 r.m.s. (km s$^{-1}$)& & &\multicolumn{3}{c}{6.2}& &\\
\hline
\end{tabular}
\end{table}

\begin{figure*}
\centering
\subfigure[He I $\lambda$ $4026$]{\label{1}\includegraphics[width=5.4cm]{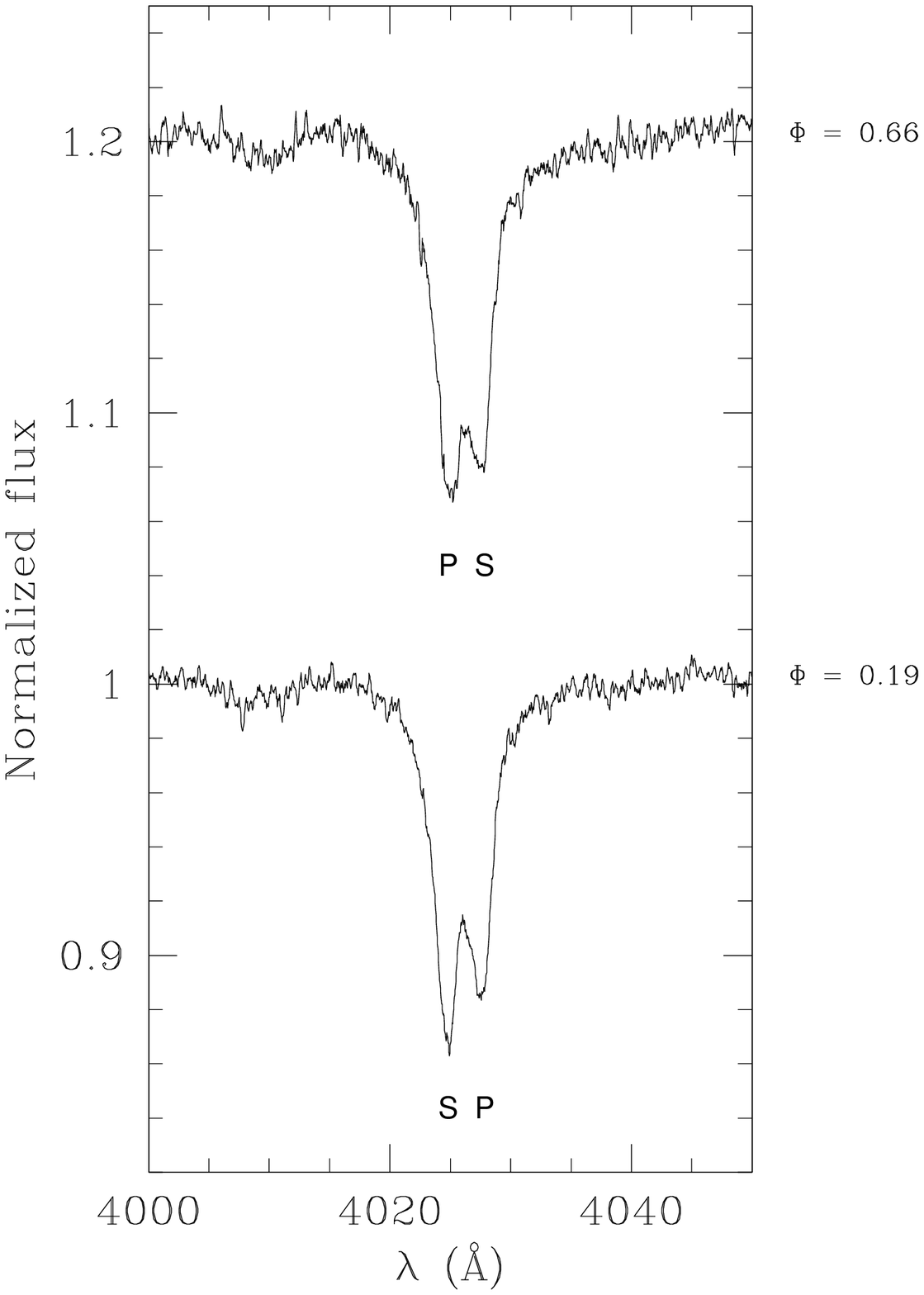}}
\subfigure[He I $\lambda$ $4471$]{\label{2}\includegraphics[width=5.4cm]{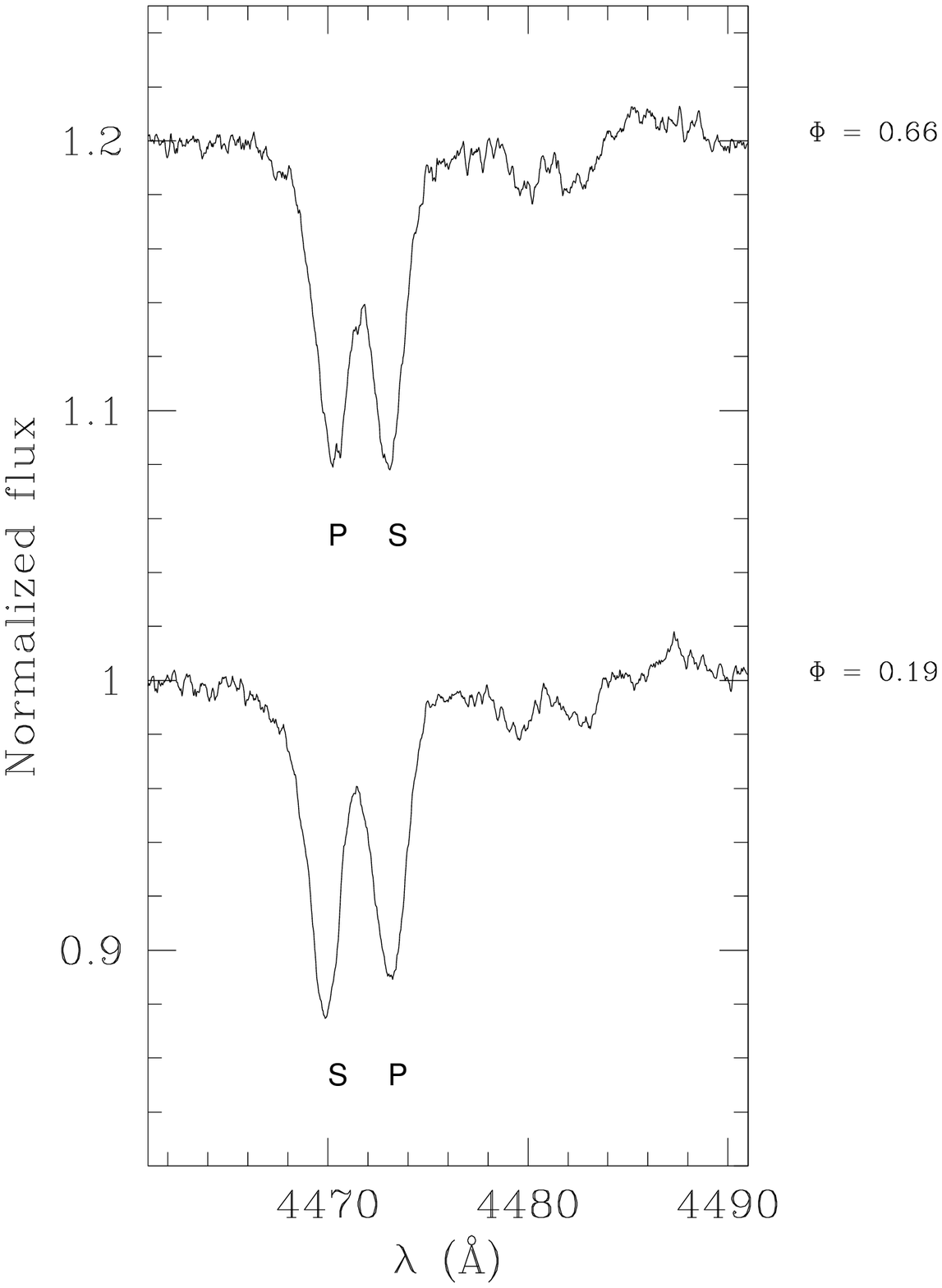}}
\subfigure[He I $\lambda$ $4921$]{\label{3}\includegraphics[width=5.4cm]{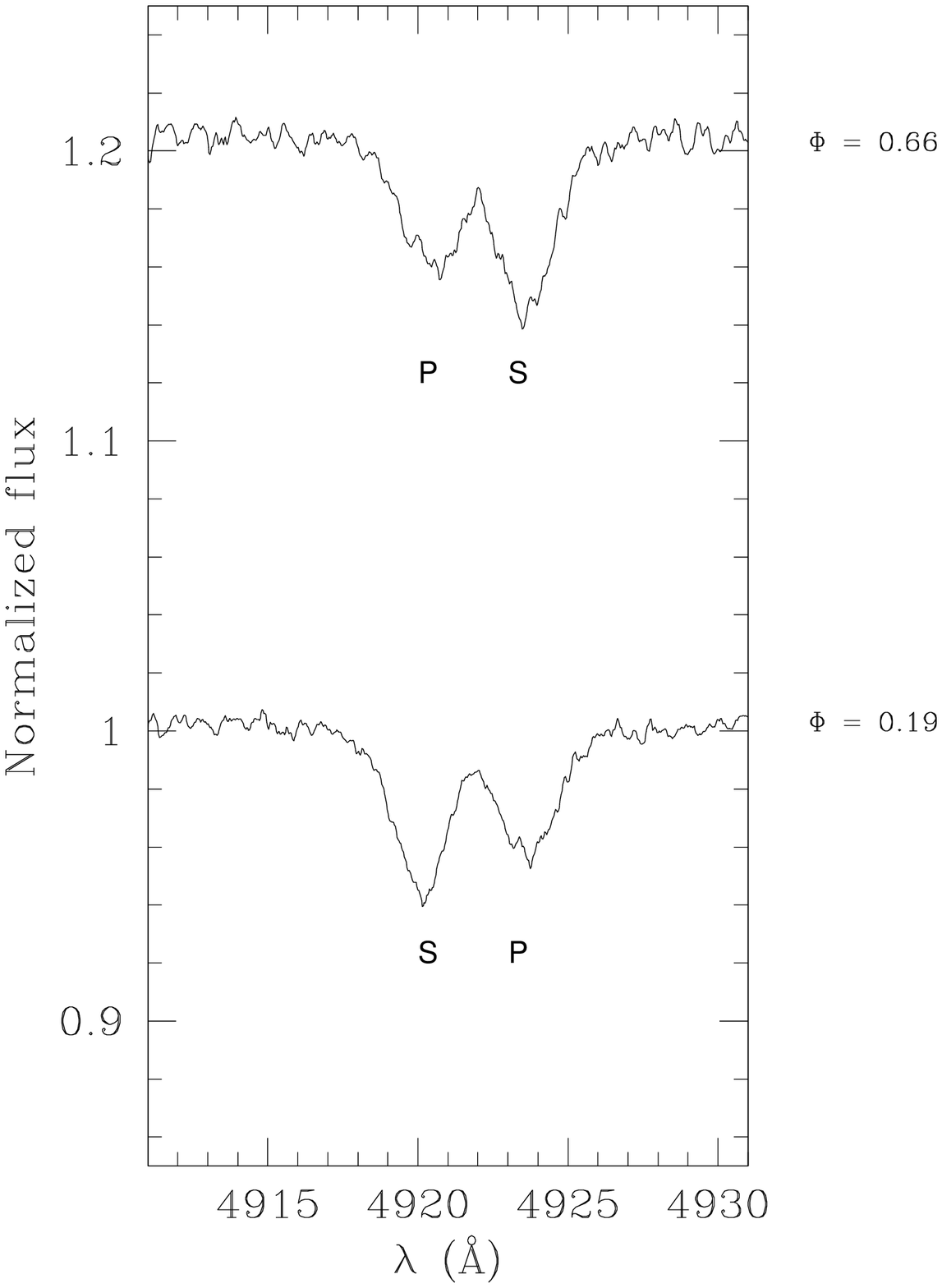}}
\caption{Different spectral lines of \object{HD~165\,052}, at opposite orbital phases: 0.19 (when the secondary is approaching) and 0.66 (when the secondary is receding). The Struve-Sahade effect is clearly visible in the He~I~$\lambda\lambda$~4026 and~4471 lines but less in He~I~$\lambda$~4921.}
\label{fig:165052ssplot}
\end{figure*}

In order to check these results, we compared the disentangled spectra of Fig.~\ref{fig:165052dis} with the spectroscopic atlas of \citet{WF90}. For the primary star, the fact that the He~I~$\lambda$~4471 and the He~II~$\lambda$~4542 lines have almost the same intensity and that the He~II~$\lambda$~4686 line is more intense than He~II~$\lambda$~4542 indicates an O7 spectral type rather than O6.5. For the secondary star, it is clear that the He~I~$\lambda$~4471 line has a greater intensity than He~II~$\lambda$~4542, which indicates a later spectral type than O7, thus O7.5 or O8. The discrepancy between the spectral type found with the method of \citet{Mathys88} and the method of \citet{WF90} seems to come from the fact that the first one is based on the equivalent width ratios, while the second one uses intensity ratios of the lines. It can be seen in Fig.~\ref{fig:165052dis} that the He~I~$\lambda$~4471 line is narrower than the He~II~$\lambda$~4542 line, which leads to smaller equivalent widths, even when the intensity is higher, and thus to earlier spectral types. Finally, the fact that the primary component of He~II~$\lambda$~4542 has an intensity larger than the secondary component in the disentangled and all observed spectra confirms that the primary star has an earlier type than the secondary star.

\subsection{Struve-Sahade effect \label{ss_165052}}

\begin{figure}
\centering
\resizebox{0.8\hsize}{!}{\includegraphics{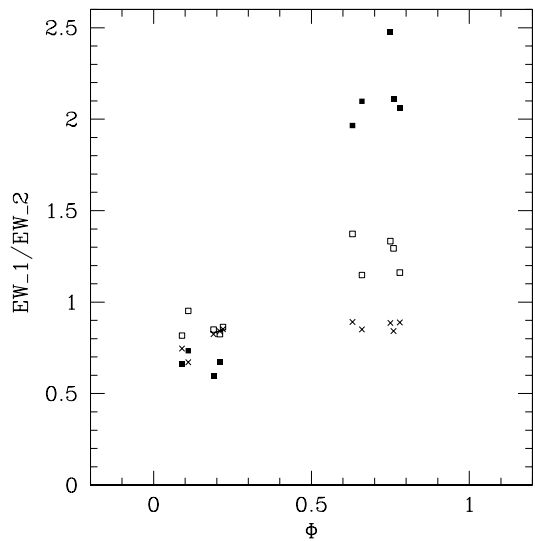}}
\caption{Equivalent width ratios (primary/secondary) of He~I~$\lambda$~4026 (filled squares), He~I~$\lambda$~4471 (open squares) and He~I~$\lambda$~4921 (crosses) in the \object{HD~165\,052} binary system, as a function of the orbital phase~$\Phi$.}
\label{fig:rapport_EW}
\end{figure}

\citet{AMB02} claimed the detection of a S-S~effect in this system for the He~I~$\lambda\lambda$~4471, 4921~lines but they did not provide quantitative results. The presence of the Struve-Sahade effect is confirmed for the He~I~$\lambda$~4471 line by our FEROS spectra. Indeed, we have measured the equivalent width of this line at different orbital phases, and its value is 21\% larger when the star is approaching compared to phases when the same star is receding, both for the secondary as well as for the primary star of the system. The same behavior appears, stronger, in He~I~$\lambda$~4026: the equivalent width of the lines of either star is 85\% larger when it is approaching than when it is receding. On the other hand, no significant effect could be detected in the He~I~$\lambda$~4921~line nor in any other line between 3900 and 7000~\AA~(see Fig.~\ref{fig:165052ssplot}). To quantify the impact of normalisation errors, the same measurements have been made on a diffuse interstellar band (DIB, $\lambda = 5780$\AA) and the level of the noise in the DIB equivalent widths is low compared to the variations detected in the He I~lines (the standard deviation of all measurements is about 2\%). Furthermore, the variations of the ratio between the equivalent width of the primary and of the secondary component have the same behavior for the two He~I~lines, as can be seen in Fig. \ref{fig:rapport_EW}, which indicates that the variations detected in He I $\lambda$~4471 are probably real and are not due to noise or normalisation errors, even if they are somewhat weaker than the variations in He~I~$\lambda$~4026. This ratio is almost constant all over the orbital cycle for the He~I~$\lambda$~4921 line. Finally, we have checked that no DIB exists on the blue side of the He~I~$\lambda$~4026 line that could bias the equivalent width measurement of the approaching component.

\section{HD 100\,213}

\object{HD~100\,213}, also called TU Mus, is an eclipsing contact binary with a very short period ($P \sim 1.39$ days). It has been studied in the optical wavelength domain both spectroscopically and photometrically by \citet{AG75} who described it as a binary system containing unevolved stars of total mass around 40 M$_{\odot}$. These authors also computed an orbital solution, with radial velocity amplitudes equal to 251.3 $\pm$ 4.8 and 370.5 $\pm$ 4.3 km s$^{-1}$, for the primary and the secondary, respectively. On the other hand, \citet{SLK95} analysed IUE data and found a large discrepancy with the radial velocities of \citet{AG75}, obtaining smaller values of $K_1 = 216.7 \pm 2.7$ and $K_2 = 345.4 \pm 3.1$ km s$^{-1}$ and thus lower masses. They attributed this fact to the rather small number of lines measured by \citet{AG75} and to their incomplete coverage of the orbital cycle. However, \citet{TMZ03} calculated a new orbital solution using nine FEROS spectra  and found results in good agreement with those of \citet{AG75}. The discrepancy between UV and optical data either reflects a genuine physical effect or could result from the methods used to evaluate the radial velocities. Indeed, \citet{SLK95} used cross-correlations with a template star (19~Cep,~O9.5~Ib) while \citet{AG75}, as well as \citet{TMZ03}, performed direct measurements on the observed spectra.

We observed this system 11 times with the FEROS echelle spectrometer. In 2002 the instrument was mounted on the 1.5m telescope (with a mean SNR equal to 340), while in 2003 it was on the 2.2m telescope (mean SNR = 330). The journal of observations is presented in Table~\ref{tab:journal:100213}. The data reduction is identical to the one of \object{HD~165\,052}.

\subsection{Orbital solution}

\begin{table}
\caption{Same as Table~\ref{tab:journal:165052} but for \object{HD~100\,213}.}
\label{tab:journal:100213}
\centering
\begin{tabular}{c c r r}
\hline
\hline
HJD &$\Phi$ &$RV_1$~~~&$RV_2$~~~\\    
$-$ 2 450 000 & &(km s$^{-1}$) & (km s$^{-1}$)\\
\hline
\multicolumn{4}{c}{}\\
2335.5871&0.10&118.6~~~& $-$276.4~~~\\
2335.7079&0.19&207.3~~~& $-$361.2~~~\\
2337.5760&0.54&$-$31.6~~~& $-$31.6~~~\\
2337.7070&0.63&$-$193.2~~~&  239.1~~~\\
2338.5718&0.25&235.7~~~& $-$384.3~~~\\
2338.7038&0.35&208.1~~~& $-$324.6~~~\\
2339.5706&0.97&$-$22.4~~~&$-$22.4~~~\\
2339.7030&0.07&3.0~~~&  3.0~~~\\
2782.5876&0.31&226.0~~~& $-$361.6~~~\\
2783.6331&0.06&13.4~~~&  13.4~~~\\
2784.6100&0.77&$-$260.8~~~&  347.9~~~\\
\hline
\end{tabular}
\end{table}

\begin{table}
\caption{Amplitudes of the radial velocity curves (in km~s$^{-1}$) of the He lines in the \object{HD~100\,213} spectra. Quoted uncertainties are the 1$\sigma$ error bars.}
\label{tab:k1k2}
\centering
\begin{tabular}{c c c}
\hline
\hline
Line&$K_1$&$K_2$\\
\hline
\multicolumn{3}{c}{}\\
He I $\lambda$ 4026  &246.5 $\pm$ 3.3&363.8 $\pm$ 4.8\\
He I $\lambda$ 4471  &244.5 $\pm$ 3.3&372.3 $\pm$ 5.1\\
He II $\lambda$ 4686 &251.3 $\pm$ 4.6&355.2 $\pm$ 6.5\\
He I $\lambda$ 4921  &245.9 $\pm$ 1.2&373.0 $\pm$ 1.9\\
He II $\lambda$ 5412 &239.8 $\pm$ 4.6&351.2 $\pm$ 6.8\\
He I $\lambda$ 5876  &269.7 $\pm$ 7.4&388.8 $\pm$ 10.7\\
\hline
\end{tabular}
\end{table}

The radial velocities were measured on the spectra by fitting two Gaussian line profiles on the data in order to separate the primary and secondary components. It was not possible to apply the disentangling method for this system, because the method failed to produce consistent results. The reason for this failure is probably that different spectral lines yield different radial velocities (see Table~\ref{tab:k1k2}). The number of available spectra was too low to allow an improvement of the orbital period and we thus fixed the period to the value found by \citet{TMZ03}. 

\begin{figure}
\centering
\resizebox{0.9\hsize}{!}{\includegraphics[width=7.1 cm]{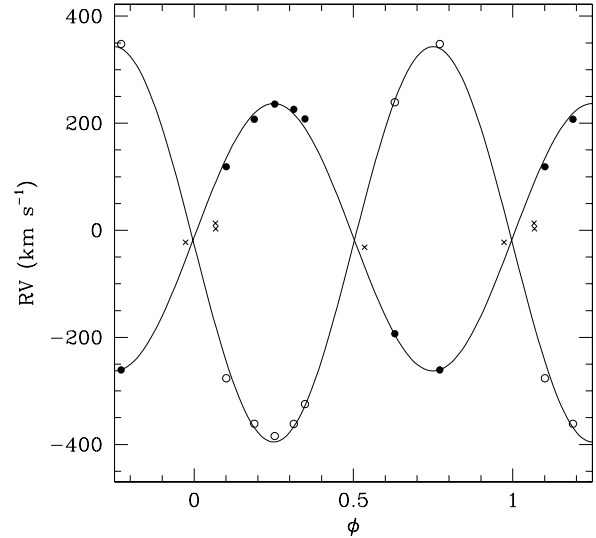}}
\caption{Radial velocity curve of \object{HD~100\,213} determined from FEROS observations and computed with LOSP. The RVs have been determined by a standard two Gaussian fitting of the lines with MIDAS (see Table~\ref{tab:journal:100213}). Primary data are represented by filled symbols, while secondary data are represented by open symbols. Crosses represent phases for which no deblending was possible. They were not included in the fit. $\Phi$~=~0 corresponds to the phase of secondary eclipse.}
\label{fig:vrad:100213}
\end{figure}

The lines used in the calculation of the orbital solution are: He~I~$\lambda$~4026, He~I~$\lambda$~4471, He~II~$\lambda$~4686, He~I~$\lambda$~4921, He~II~$\lambda$~5412 and He~I~$\lambda$~5876. Different lines yield different orbital solutions (see Table~\ref{tab:k1k2}). This is especially the case when comparing the values of $K_2$. Although we did not find a clear trend with the excitation potential of the atomic levels of the lines, we see that He~II lines yield lower variations of the radial velocity than the He~I lines. Fig. \ref{fig:vrad:100213} shows the radial velocity curve of \object{HD~100\,213} as a function of the orbital phase. Table \ref{tab:orbit:100213} gives the orbital solution based on the mean RVs of all lines computed with the LOSP program. This solution is close to the one found by \citet{AG75} and \citet{TMZ03}, with $K_1$ and $K_2$ equal to 250 and 369 km s$^{-1}$. It is therefore possible that the UV lines also have different radial velocities because they form over different parts of the stellar surface (see below), hence explaining the difference between the IUE and optical orbital solutions.

\subsection{Spectral type determination \label{spectraltype100213}}

\begin{table}
\caption{Same as Table \ref{tab:orbit:165052} but for \object{HD~100\,213}.}
\label{tab:orbit:100213}
\centering
\begin{tabular}{lrclcrcl}
\hline
\hline
Parameter           &\multicolumn{3}{c}{Primary}& &\multicolumn{3}{c}{Secondary}\\
\hline
\multicolumn{8}{c}{}\\
 $P$ (days)& &  &\multicolumn{3}{c}{1.3873 (fixed)}& & \\
 $e$ & & &\multicolumn{3}{c}{0 (fixed)}& & \\
 $T_0$ (HJD)& & &  \multicolumn{3}{c}{2452334.059 $\pm$ 0.005}& & \\
 $\gamma$ (km s$^{-1}$)&   \multicolumn{3}{c}{$-$13.2 $\pm$ 5.9}& &\multicolumn{3}{c}{$-$26.3 $\pm$ 6.9} \\
 $K$ (km s$^{-1}$)&  \multicolumn{3}{c}{249.7 $\pm$ 6.2}& &\multicolumn{3}{c}{369.2 $\pm$ 9.1}\\
 $a$ $sin i$ ($R_{\odot}$)&  \multicolumn{3}{c}{6.84 $\pm$ 0.17}& &\multicolumn{3}{c}{10.12 $\pm$ 0.25} \\
 $M$ $sin^3 i$ ($M_{\odot}$) &    \multicolumn{3}{c}{20.32 $\pm$ 1.07}& &\multicolumn{3}{c}{13.74 $\pm$ 0.65}\\
 $Q$ ($  M1/M2 $) & & & \multicolumn{3}{c}{1.48 $\pm$ 0.06}& &\\ 
 r.m.s. (km s$^{-1}$)& & &\multicolumn{3}{c}{13.6}& &\\
\hline
\end{tabular}
\end{table}

We have determined the spectral type of \object{HD~100\,213} following Conti's criteria, as for \object{HD~165\,052}, and found O7.5 and O9.5 for the primary and the secondary star, respectively. The He~II~$\lambda$~4686~line is also in absorption, for both stars. When the optical brightness ratio of the stars is taken into account ($\frac{L_1}{L_2}\,=\,1.88$, determined in the same way as for \object{HD~165\,052}), measurements of the equivalent widths of this line give  EW$_{\lambda}(4686)\,=\,0.78$~\AA~for the primary star, which indicates that it is a main sequence star. Furthermore, Mathys' O8.5-O9.7 luminosity criterium ($\log$~W$_{\lambda}'''\,= \,\log$~EW$_{\lambda}(4388) \,+\,\log$~EW$_{\lambda}(4686)$) yields the same result for the secondary star, with $\log$~W'''~$=\,5.43\,>\,5.40$ where the EWs are expressed in~m\AA~\citep{Mathys88}.

\begin{figure}
\centering
\subfigure[He~I~$\lambda$~4026]{\resizebox{0.8\hsize}{!}{\label{fig:ss:100213:4026}\includegraphics{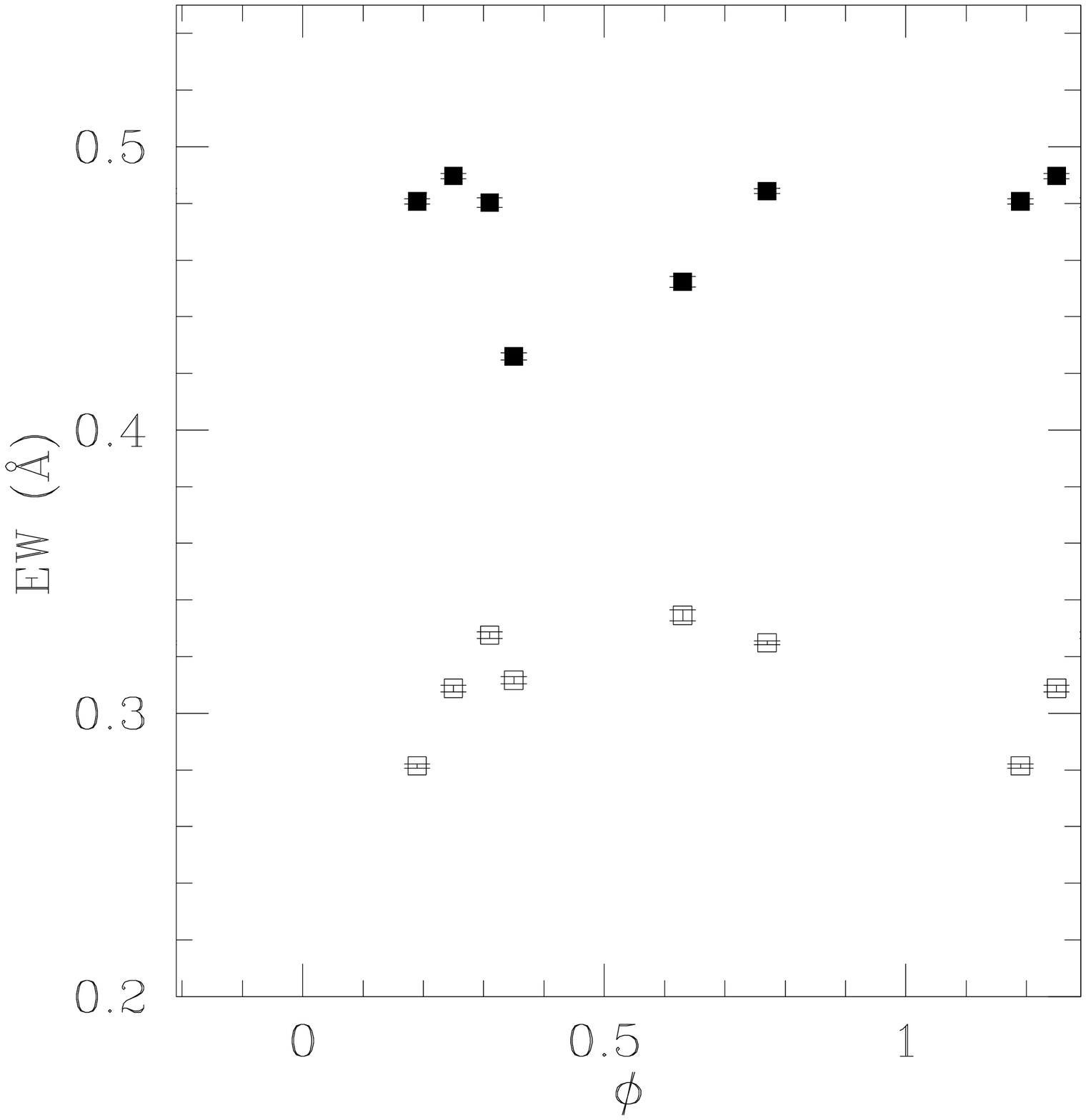}}}
\subfigure[He~I~$\lambda$~4471]{\resizebox{0.8\hsize}{!}{\label{fig:ss:100213:4471}\includegraphics{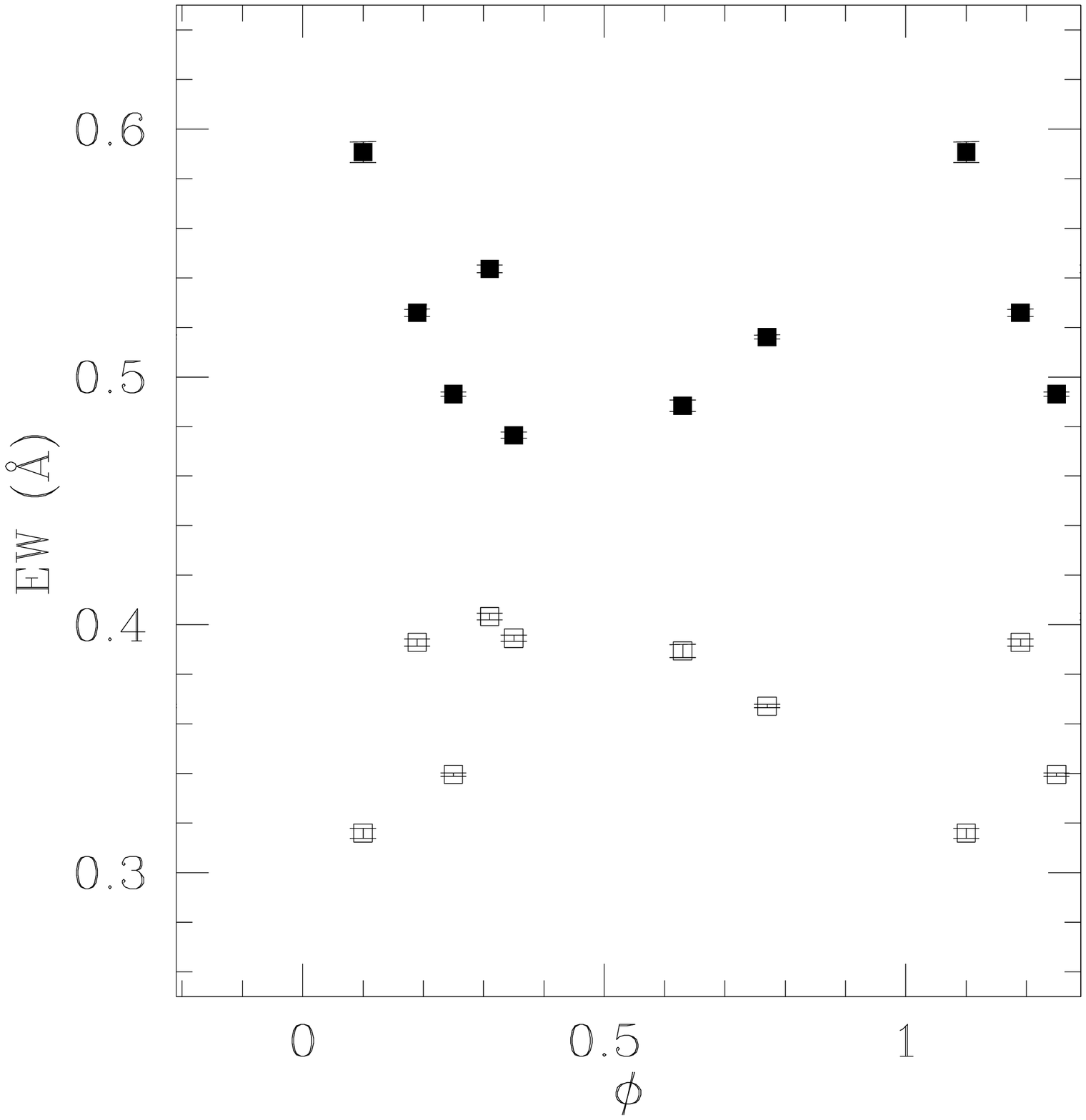}}}
\caption{Equivalent widths (in \AA) of two helium lines in the spectrum of \object{HD~100\,213}, as a function of phase. Filled and open squares represent primary and secondary data, respectively. $\Phi = 0$ corresponds to the phase of secondary eclipse. The error bars are formal errors of the Gaussian fits to the line profiles.}
\label{fig:ss:100213}
\end{figure}

\subsection{Photometry}

We performed an analysis of the Hipparcos photometry of \object{HD~100\,213} with the {\itshape Nightfall} program \footnote{For details see the Nightfall User Manual by Wichmann (1998) available at the URL: http://www.hs.uni-hamburg.de/DE/Ins/Per/Wichmann/Nightfall.html.}.This code is based on a generalised Wilson-Devinney method assuming a standard Roche geometry. In the fit, we restricted the number of free parameters to a minimum. The orbital parameters were adopted from Table~\ref{tab:orbit:100213}. For the spectral types found in Sect.~\ref{spectraltype100213}, \citet{MSH05} give values of the effective temperatures of 35100 and 31500 K for the primary and the secondary, respectively. These temperatures  did not vary significantly during the fit. When taking into account the `third light' due to a faint source to the south-east of the system (at a distance of 16 arcseconds, this source has a visual magnitude equal to 13.27), our results are similar to those of \citet{TMZ03}. We find that \object{HD~100\,213} is an overcontact binary with an inclination equal to 77.8\degr, which leads to a primary mass of 21.7~M$_{\odot}$ and a secondary mass of 14.7~M$_{\odot}$.

\subsection{Struve-Sahade effect}

In their study of OB stars from IUE spectroscopy, \citet{HSH97} listed \object{HD~100\,213} as a binary system displaying the S-S effect. Unfortunately, these authors did not provide quantitative results and did not discuss any line in particular. 

Our data confirm the variability of the EWs of the absorption lines, but the situation looks more complicated than expected for the ''classical'' Struve-Sahade effect. Fig. \ref{fig:ss:100213} shows the equivalent width of two different lines as a function of phase. Both He~I~$\lambda\lambda$~4026 and 4471 display variations of their equivalent width that are phase-locked with the orbital motion. For both stars, the EWs are maximum when the star is in front and minimum half a period later. This trend is especially well marked for the He~I~$\lambda$~4471 line. Since \object{HD~100\,213} is an eclipsing binary, this could partially be due to the mutual occultation of the two components. We used {\itshape Nightfall} to evaluate the light variations due to the ellipticity of both stars. We have then used this result to calculate the theoretical variations in equivalent widths that are exclusively due to the eclipses. Comparing this curve with the observations, we can see that all observation points of Fig. \ref{fig:ss:100213} fall outside the eclipses and occultations thus cannot be the cause of the variations. The most probable scenario is that the He~I lines are not formed homogeneously over the entire surface of the star but preferentially on the rear side situated on the opposite side of the companion. Such a scenario would provide variations similar to those seen in Fig. \ref{fig:ss:100213}.

The other lines of this system, in the studied wavelength domain, do not show any clear-cut variations, except perhaps He~I~$\lambda$~5876. However, the large number of telluric lines that affect this line renders the measurement of its equivalent width too uncertain. It seems that at least the secondary star shows the same variations as the He~I~$\lambda\lambda$~4026 and 4471 lines. Thus the He~I~$\lambda$~5876 line should also be formed preferentially over the back side of the stars, and this fact seems to be confirmed by the large radial velocity amplitude associated with it (see Table~\ref{tab:k1k2}).

As for \object{HD~165\,052}, the deep interstellar lines do not vary significantly, with a standard deviation of the EW values equal to 0.7\%~while $\sigma$ is of the order of 3\% for the two helium lines considered in Fig. \ref{fig:ss:100213}. Furthermore, there are no DIBs blended with any of the measured lines.

\section{HD~159\,176}

\object{HD~159\,176} is a double line spectroscopic, non eclipsing binary \citep{Trumpler30} situated in the young open cluster NGC~6383 and embedded in a large H~II region. \citet{CCJ75} derived an orbital solution from optical spectra. They adopted a null eccentricity and found a period of 3.366 days and an inclination of 50\degr \, or less. Following the \citet{CA71} system, they determined spectral types equal to O7~V~+~O7~V. \citet{Thomas75} obtained photometric data of this system and found ellipsoidal variations but no eclipse. The photometric solution indicated a bolometric magnitude equal to $-$8.6 and effective temperatures equal to 38000 K. More recently, \citet{SKP93} derived an orbital solution from IUE data, also assuming a zero eccentricity. Their period, equal to 3.367 days, is almost the same as the period found by \citet{CCJ75}.  Their mass ratio is very close to 1, which means equal spectral types for both components. On the other hand, \citet{Pachoulakis96} studied \object{HD~159\,176} with UV, optical and photometric data and found that the primary star is actually hotter than the secondary star and has thus an earlier type, suggesting an O6~V~+~O7~V binary. An XMM-Newton study of \object{HD~159\,176} revealed an excess of X-ray luminosity by a factor $\sim$7 compared to the expected value for a single O-star \citep{DBRP04}, which indicates the presence of a wind-wind interaction in the system. 

\begin{table}
\caption{Same as Table~\ref{tab:journal:165052} but for \object{HD~159\,176}.}
\label{tab:journal:159176}
\centering
\begin{tabular}{c c r r}
\hline
\hline
HJD &$\Phi$ &$RV_1$~~~&$RV_2$~~~\\    
$-$ 2 450 000 & &(km s$^{-1}$) & (km s$^{-1}$)\\
\hline
\multicolumn{4}{c}{}\\
1299.9035&0.41&138.8~~~& $-$101.2~~~\\
1300.9147&0.71&$-$189.7~~~&  211.7~~~\\
1300.9191&0.71&$-$198.6~~~&  220.2~~~\\
1301.9202&0.00&18.0~~~&    17.4~~~\\
1302.9315&0.30&191.4~~~& $-$198.5~~~\\
1304.9234&0.90&$-$119.9~~~& 132.0~~~\\
1327.7738&0.68&$-$179.6~~~&  208.7~~~\\
2037.7203&0.55&$-$69.3~~~&   61.5~~~\\
2335.8786&0.11&150.2~~~& $-$131.0~~~\\
2336.8672&0.41&138.2~~~& $-$109.7~~~\\
2337.8773&0.71&$-$190.1~~~&  219.4~~~\\
2338.8678&0.00&9.4~~~&    0.6~~~\\
2338.8932&0.01&39.0~~~&    $-$27.5~~~\\
2339.8702&0.30&200.5~~~& $-$199.9~~~\\
2381.8554&0.77&$-$199.6~~~&  229.3~~~\\
2382.8421&0.06&108.0~~~&    $-$60.8~~~\\
2383.8469&0.36&160.6~~~& $-$167.9~~~\\
2782.7632&0.85&$-$151.4~~~&  198.1~~~\\
2783.7571&0.15&188.2~~~& $-$151.9~~~\\
2784.7995&0.46&79.2~~~&   $-$50.0~~~\\
\hline
\end{tabular}
\end{table}

\begin{table}
\caption{Same as Table~\ref{tab:orbit:165052} but for \object{HD~159\,176}.}
\label{tab:orbit:159176}
\centering
\begin{tabular}{lrclcrcl}
\hline
\hline
 Parameter  &\multicolumn{3}{c}{Primary}& &\multicolumn{3}{c}{Secondary}\\
\hline
\multicolumn{8}{c}{}\\
 $P$ (days)& &  &\multicolumn{3}{c}{3.36673  $\pm$  0.00004}& & \\
 $e$ & & &\multicolumn{3}{c}{0 (fixed)}& & \\
 $T_0$ (HJD)& & &  \multicolumn{3}{c}{2451298.539   $\pm$ 0.006}& & \\
 $\gamma$ (km s$^{-1}$)&   \multicolumn{3}{c}{12.1 $\pm$ 2.5}& &\multicolumn{3}{c}{8.9 $\pm$ 2.5} \\
 $K$ (km s$^{-1}$)&  \multicolumn{3}{c}{210.3 $\pm$ 3.5}& &\multicolumn{3}{c}{216.5 $\pm$ 3.6}\\
 $a$ $sin i$ ($R_{\odot}$)&  \multicolumn{3}{c}{14.0 $\pm$ 0.2}& &\multicolumn{3}{c}{14.4 $\pm$ 0.2} \\
 $M$ $sin^3 i$ ($M_{\odot}$) &    \multicolumn{3}{c}{13.7 $\pm$ 0.5}& &\multicolumn{3}{c}{13.3 $\pm$ 0.5}\\
 $Q$ ($  M1/M2 $) & & & \multicolumn{3}{c}{1.03 $\pm$ 0.03}& &\\
 r.m.s. (km s$^{-1}$)& & &\multicolumn{3}{c}{10.3}& &\\
\hline
\end{tabular}
\end{table}

The observations of this system were taken between 1999 and 2003, with the FEROS spectrometer (20 spectra). The journal of observations is presented in Table~\ref{tab:journal:159176}. As for \object{HD~100\,213}, the instrument was mounted on the 1.5m telescope in 1999, 2001 and 2002 (mean SNR = 350), while in 2003 it was on the 2.2m (mean SNR = 430). The data reduction was the same as for \object{HD~100\,213} and \object{HD~165\,052}.

\subsection{Orbital solution}

\begin{figure}
\centering
\resizebox{\hsize}{!}{\includegraphics{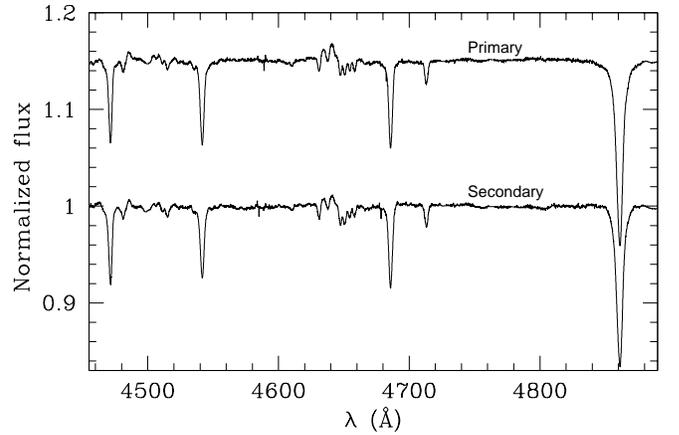}}
\caption{Normalised disentangled spectra of \object{HD~159\,176}, between 4455 and 4895 \AA. The primary spectrum was vertically shifted by 0.15 units for clarity.}
\label{fig:159176dis}
\end{figure}

\begin{figure}
\centering
\resizebox{0.9\hsize}{!}{\includegraphics[width=7.1 cm]{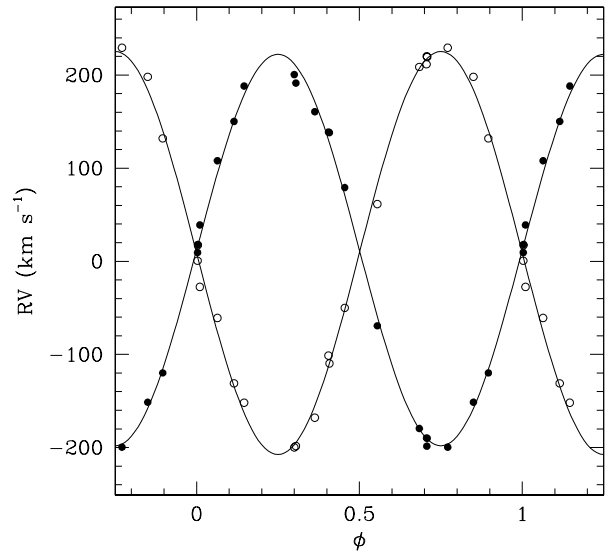}}
\caption{Radial velocity curve of \object{HD~159\,176} determined from FEROS observations and computed with LOSP. The RVs were determined through a disentangling of the primary and secondary spectra. Primary data are represented by filled symbols, secondary data are represented by open symbols. $\Phi = 0$ corresponds to the phase of conjunction with the secondary star behind.}
\label{fig:vrad:159176}
\end{figure}

We have used the disentangling method to determine the RVs that yield the orbital solution of Table~\ref{tab:orbit:159176}. The lines considered in the RV calculation are: H$_{\gamma}$, He~I~$\lambda$~4471, Mg~II~$\lambda$~4481, He~II~$\lambda$~4542, He~II~$\lambda$~4686, He~I~$\lambda$~4713 and H$_{\beta}$. The period was allowed to vary from an initial value of 3.366767 days \citep{SKP93}. Fig.~\ref{fig:159176dis} shows the disentangled spectra of \object{HD~159\,176}, and Fig.~\ref{fig:vrad:159176} the radial velocity curve as a function of the orbital phase.

\begin{figure*}
  \begin{center}
     \subfigure[He I $\lambda$ 4026]{\label{subfig:hd1591764026comp}\includegraphics[width=5.4 cm]{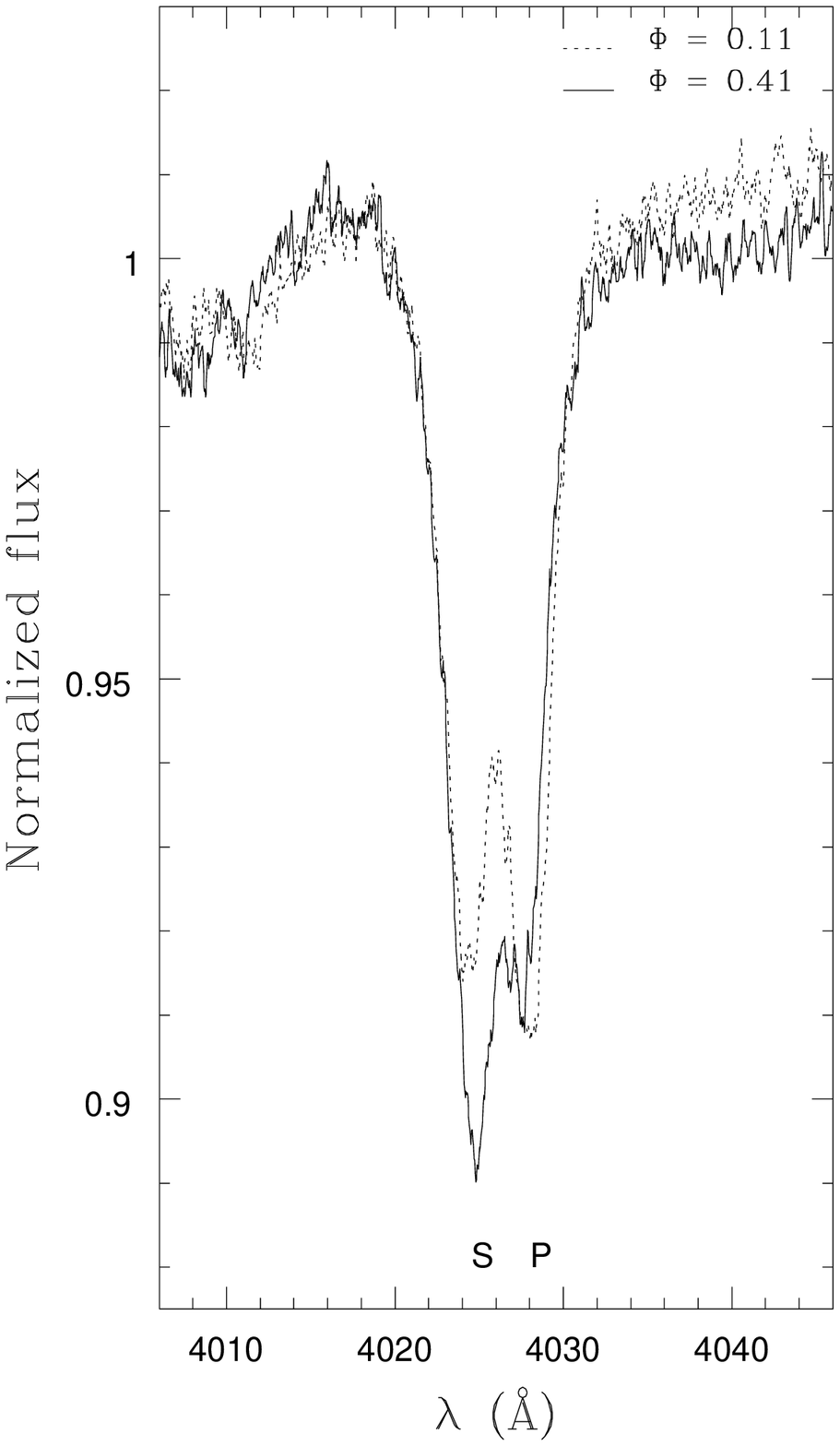}}
     \subfigure[He I $\lambda$ 4471]{\label{subfig:hd1591764471comp}\includegraphics[width=5.4 cm]{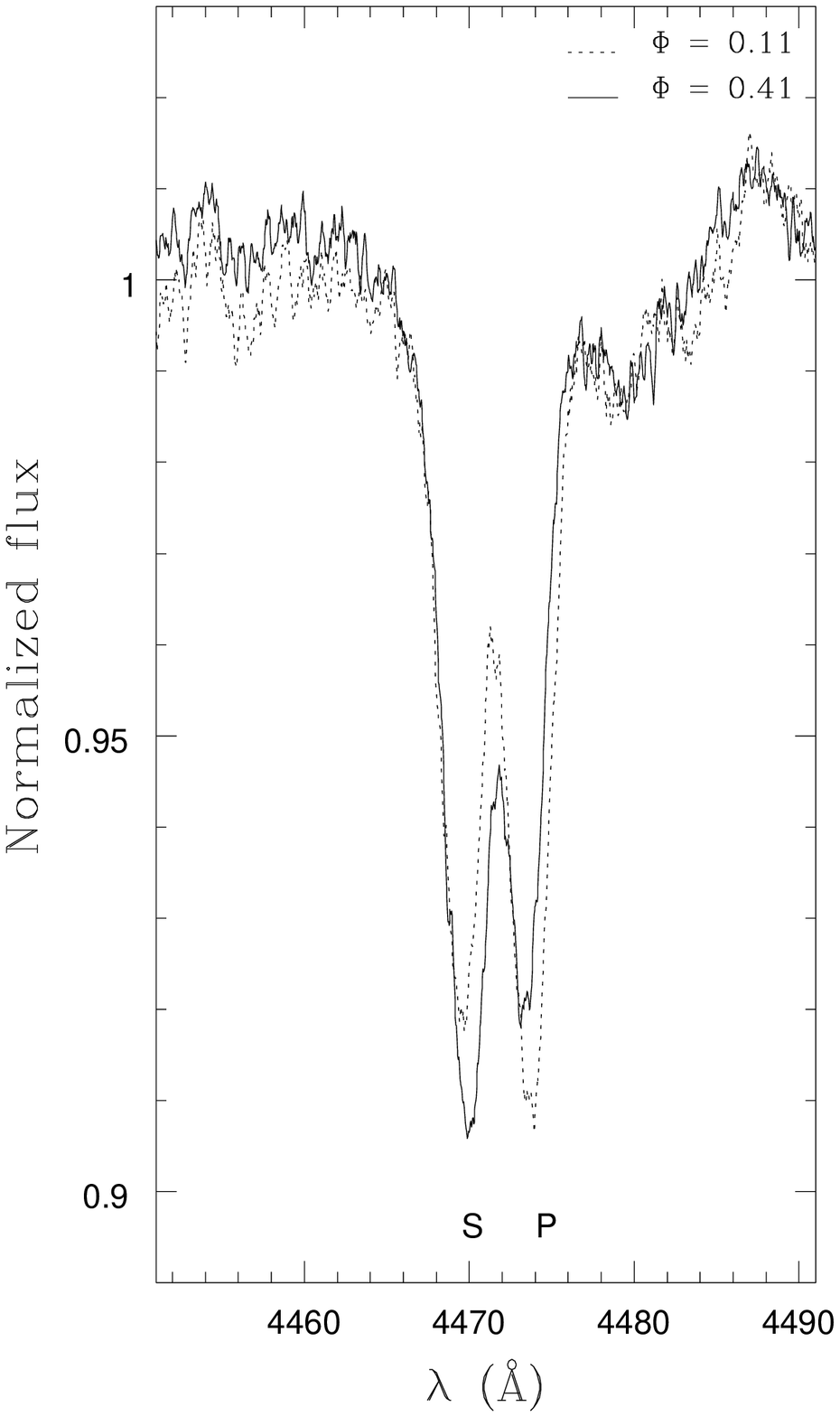}}
     \subfigure[He I $\lambda$ 4921]{\label{subfig:hd1591764921comp}\includegraphics[width=5.4 cm]{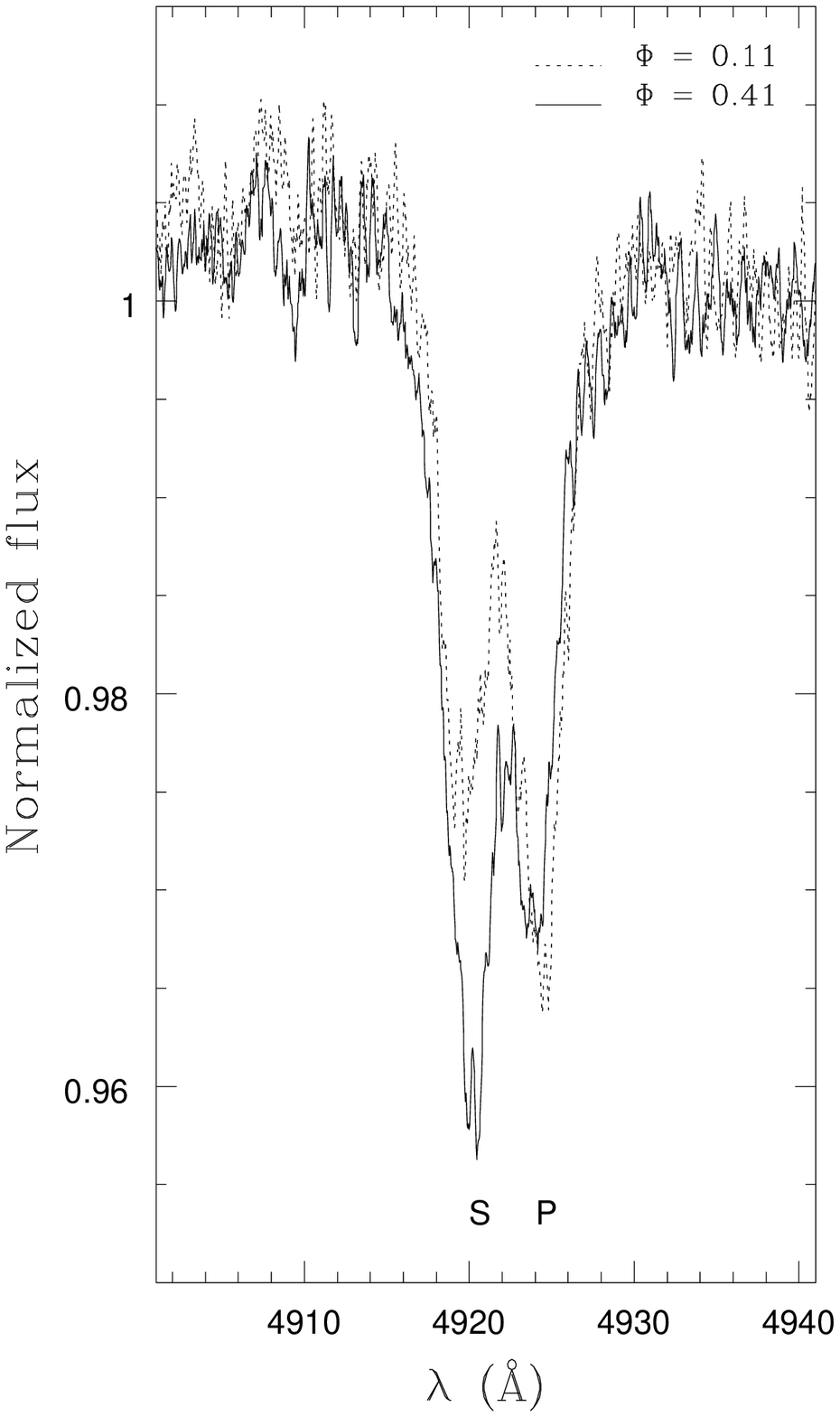}}
  \end{center}
  \caption{Comparison of three He I lines of \object{HD~159\,176} for two different orbital phases when the primary is receding.}
  \label{fig:159176comp}
\end{figure*}

\begin{figure}
\centering
\resizebox{0.8\hsize}{!}{\includegraphics{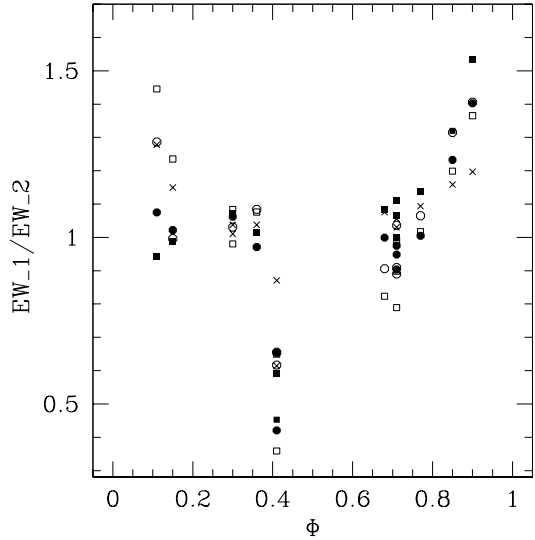}}
\caption{Equivalent widths ratio of He~I~$\lambda$~4026 (filled squares), He~I~$\lambda$~4921 (open squares), He~I~$\lambda$~4471 (filled circles), He~I~$\lambda$~4713 (open circles) and He~I~$\lambda$~5876 (crosses) in the \object{HD~159\,176} binary system, as a function of the orbital phase $\Phi$.}
\label{fig:rapport_EW2}
\end{figure}

Using the same approach as for \object{HD~165\,052}, we have determined rotational velocities of $101 \pm 5$ and $98 \pm 5$~km~s$^{-1}$ for the primary and secondary component of \object{HD~159\,176} respectively, which indicate a synchronous rotation of the system. Furthermore, the comparison between the angular velocity of the system and the angular velocity of each component yields equal values if the inclination is equal to 46\degr. This value is comparable to the 48\degr~found by \citet{SKP93} and the 50\degr~of \citet{Pachoulakis96}, which indicates that the system is probably in co-rotation. The fits of the radial velocity curve with the eccentricity as a free parameter gave $e$~=~0.02, with no improvement of the residuals, and the circularisation time calculated from \citet{Tassoul1990} is shorter than 57\,000 years. The orbit is thus considered circular.

\subsection{Spectral type determination}

The equivalent width measurements made on our individual FEROS spectra lead to spectral types of the components of \object{HD~159\,176} equal to O7~+~O7.5 on average, but uncertainties on the EWs due to the Struve-Sahade effect (see below) allow us to have a primary between  O6.5 and O7.5 and a secondary between O7 and O8. If instead we measure the EWs on the disentangled spectra, we obtain an O7 spectral type for both components. The components are main-sequence stars since the He~II~$\lambda$~4686~line is in absorption, and  EW$_{\lambda}$(4686)=~0.60 for the primary and 0.71 for the secondary star (accounting for $\frac{L_1}{L_2} = 1.18$, determined in the same manner as for \object{HD~165\,052}). The comparison of Fig.~\ref{fig:159176dis} with the spectra of \citet{WF90} confirms that both spectral types are very similar and should be very close to O7, even if the primary star seems slightly earlier. As for \object{HD~165\,052}, it can be seen in Fig.~\ref{fig:159176dis} that the N~III~$\lambda$$\lambda$~4634,4640,4641 lines are in emission, which means that both components have an O((f)) spectral type.

\subsection{Struve-Sahade effect \label{ss159176}}

The first report of the S-S effect for this system goes back to \citet{Trumpler30}, who noted a change in the appearance of the optical spectral lines. This was confirmed by \citet{CCJ75} who specified that the approaching star usually appears to have deeper lines. \object{HD~159\,176} is also cited, for the same reason, by \citet{SKP93} and \citet{HSH97}. Whilst these authors generally specify from which star and phase they have detected a strange behavior, they do not show any figures or quantitative results. 
\begin{table}
\caption{Same as Table \ref{tab:journal:165052} but for \object{DH~Cep}.}
\label{tab:journal:dhcep}
\centering
\begin{tabular}{c c r r}
\hline
\hline
HJD &$\Phi$ &$RV_1$~~~&$RV_2$~~~\\    
$-$ 2 450 000 & &(km s$^{-1}$) & (km s$^{-1}$)\\
 \hline
\multicolumn{4}{c}{}\\
2518.4440&0.70&$-$250.7~~~& 208.9~~~\\
2520.4080&0.63 &$-$210.1~~~&157.8~~~\\
2522.3963&0.57 &$-$159.3~~~&41.4~~~\\
2524.3893&0.51 &$-$69.7~~~&$-$32.0~~~\\
2528.3498&0.39 &71.6~~~&$-$228.7~~~\\
2529.3596&0.87 &$-$219.9~~~&159.1~~~\\
2532.3708&0.29 &170.3~~~&$-$300.1~~~\\
2533.3813&0.77 &$-$261.4~~~& 209.6~~~\\
2922.4670&0.09 &80.1~~~&$-$180.1~~~\\
2923.4440&0.55 &$-$100.9~~~&76.1~~~\\
3980.5385&0.32 &158.9~~~& $-$270.9~~~\\
3981.4924&0.78 &$-$298.1~~~& 208.6~~~\\
3982.3843&0.19 &151.8~~~& $-$291.8~~~\\
3982.5795&0.29 &161.3~~~& $-$320.0~~~\\
3983.5053&0.73 &$-$290.3~~~& 201.3~~~\\
3984.4928&0.19 & 159.7~~~& $-$299.7~~~\\
\hline
\end{tabular}
\end{table}

\begin{figure}
\centering
\resizebox{\hsize}{!}{\includegraphics{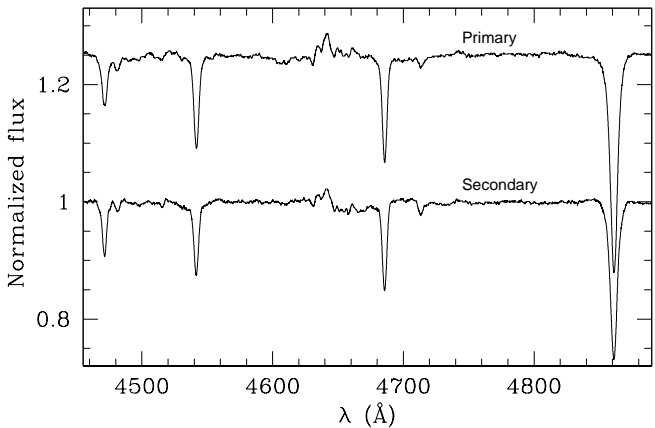}}
\caption{Normalised disentangled spectra of \object{DH~Cep}, between 4455 and 4895 \AA. The primary spectrum was vertically shifted by 0.25 units for clarity.}
\label{fig:dhcepdis}
\end{figure}
On the basis of our FEROS observations, we could observe that some variations in the line profiles appear when the secondary star is approaching (from $\Phi=0.0$ to $\Phi=0.5$). For two He~I lines measured in our wavelength domain (He~I~$\lambda$~4026 and He~I~$\lambda$~4921), there is an increase of the equivalent width and depth of the secondary star's lines  at phase $\Phi= 0.41$ compared to its value at phase $\Phi = 0.11$ (see Figs.~\ref{subfig:hd1591764026comp} and~\ref{subfig:hd1591764921comp}). Furthermore, for the He I $\lambda$$\lambda$~4471, 4713 and 5876 lines, an inversion of the relative intensity of the two component lines occurs between these two orbital phases (see Fig.~\ref{subfig:hd1591764471comp}). These changes are both due to a modification of the line maximum intensities and of the full width at half minimum. These variations are well above the level of noise, measured from the variations of the diffuse interstellar band at 5780~\AA~($\sigma = 0.8$\%) and no DIB exists in any measured lines that could modify the measurements. Fig.~\ref{fig:rapport_EW2} shows that the ratio of primary and secondary equivalent widths of all these considered lines also show the same behavior with a decrease of the ratio at phase~0.41. This decrease occurs in both available spectra at this phase, one in~1999 and the other one in~2002. When the primary star is approaching, the EW ratio increase is due to a progressive increase of the EW of the primary star, while the secondary equivalent width is decreasing. The general trend in Fig.~\ref{fig:rapport_EW2} is thus that the EWs of the He~I lines of a given component tend to be larger when this component is in front. While this effect cannot be due to occultations (\object{HD~159\,176} does not display eclipses), it could indicate that the He~I lines preferentially form on the rear sides of the stars. 

\section{DH~Cep}

\begin{table}
\caption{Same as Table~\ref{tab:orbit:165052} but for \object{DH~Cep}.}
\label{tab:orbit:dhcep}
\centering
\begin{tabular}{lrclcrcl}
\hline
\hline
 Parameter &\multicolumn{3}{c}{Primary}& &\multicolumn{3}{c}{Secondary}\\
\hline
\multicolumn{8}{c}{}\\
 $P$ (days)& &  &\multicolumn{3}{c}{2.11095  $\pm$  0.00013}& & \\
 $e$ & & &\multicolumn{3}{c}{0 (fixed)}& & \\
 $T_0$ (HJD)& & &  \multicolumn{3}{c}{2452516.976   $\pm$ 0.006}& & \\
 $\gamma$ (km s$^{-1}$)&   \multicolumn{3}{c}{-52.7 $\pm$ 4.4}& &\multicolumn{3}{c}{-45.9 $\pm$ 4.7} \\
 $K$ (km s$^{-1}$)&  \multicolumn{3}{c}{225.0 $\pm$ 4.8}& &\multicolumn{3}{c}{265.1 $\pm$ 5.7}\\
 $a$ $sin i$ ($R_{\odot}$)&  \multicolumn{3}{c}{9.4 $\pm$ 0.2}& &\multicolumn{3}{c}{11.1 $\pm$ 0.2} \\
 $M$ $sin^3 i$ ($M_{\odot}$) &    \multicolumn{3}{c}{13.9 $\pm$ 0.6}& &\multicolumn{3}{c}{17.8 $\pm$ 0.5}\\
 $Q$ ($  M1/M2 $) & & & \multicolumn{3}{c}{1.18 $\pm$ 0.04}& &\\
 r.m.s. (km s$^{-1}$)& & &\multicolumn{3}{c}{14.2}& &\\
\hline
\end{tabular}
\end{table}

\begin{figure}
\centering
\resizebox{0.9\hsize}{!}{\includegraphics{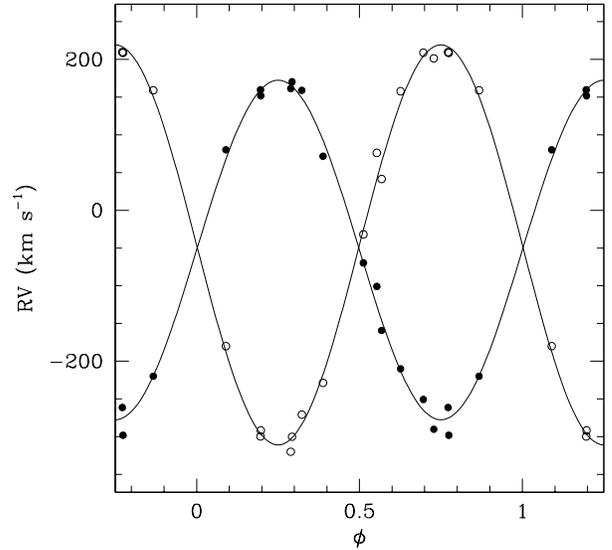}}
\caption{Radial velocity curve of \object{DH~Cep} determined from Aur\'elie observations and computed with LOSP. The RVs have been determined through a disentangling of the primary and secondary spectra. Primary data are represented by filled symbols, while secondary data are represented by open symbols. $\Phi = 0$ corresponds to the phase of the phase of conjunction with the secondary star behind.}
\label{fig:vrad:dhcep}
\end{figure}

\begin{figure*}
  \begin{center}
     \subfigure[He II $\lambda$ 4542]{\label{subfig:dhcep4542}\includegraphics[width=5.4 cm]{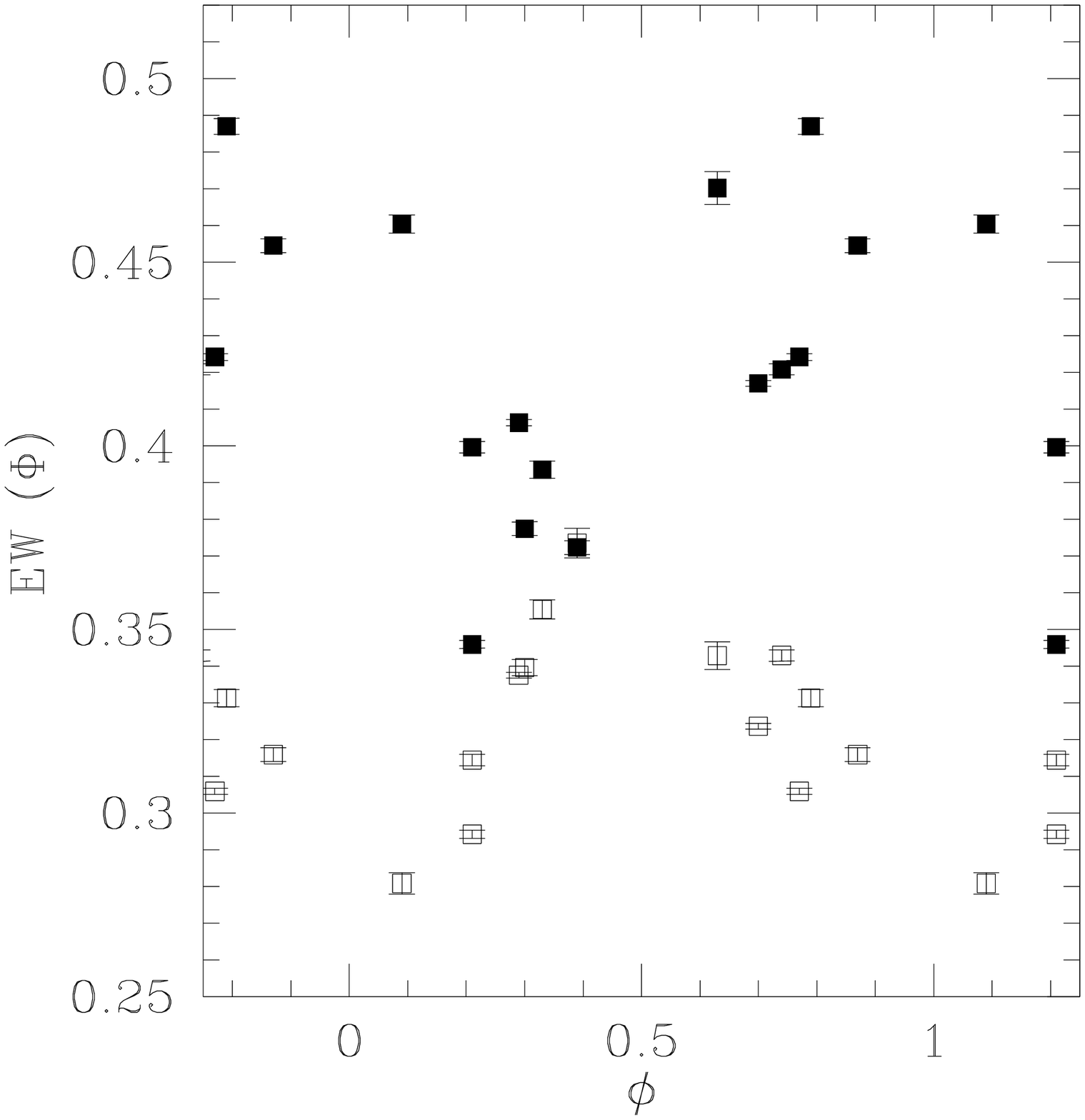}}
     \subfigure[He II $\lambda$ 4686]{\label{subfig:dhcep686}\includegraphics[width=5.4 cm]{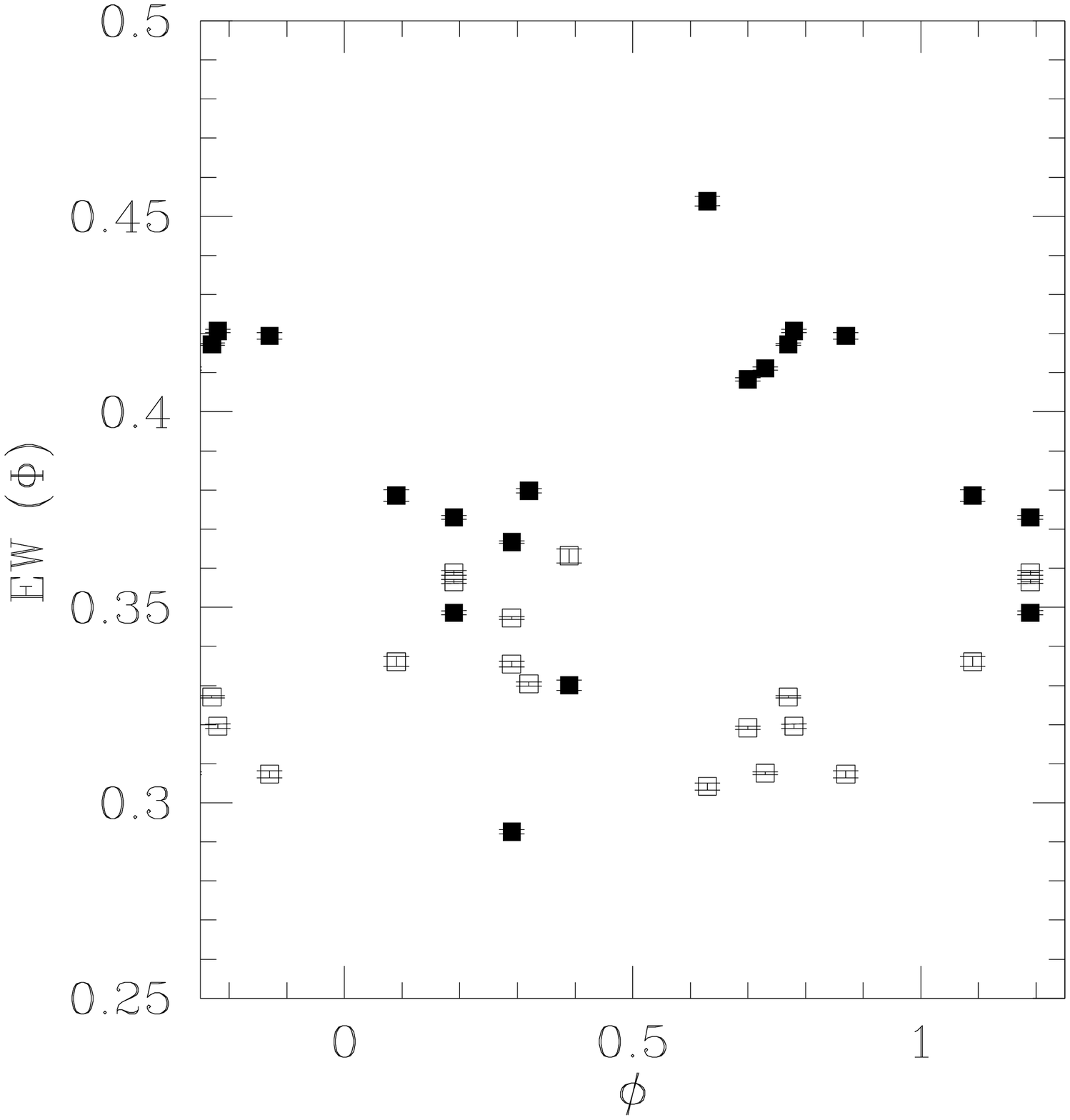}}
     \subfigure[H$_{\beta}$]{\label{subfig:dhcephbeta}\includegraphics[width=5.4 cm]{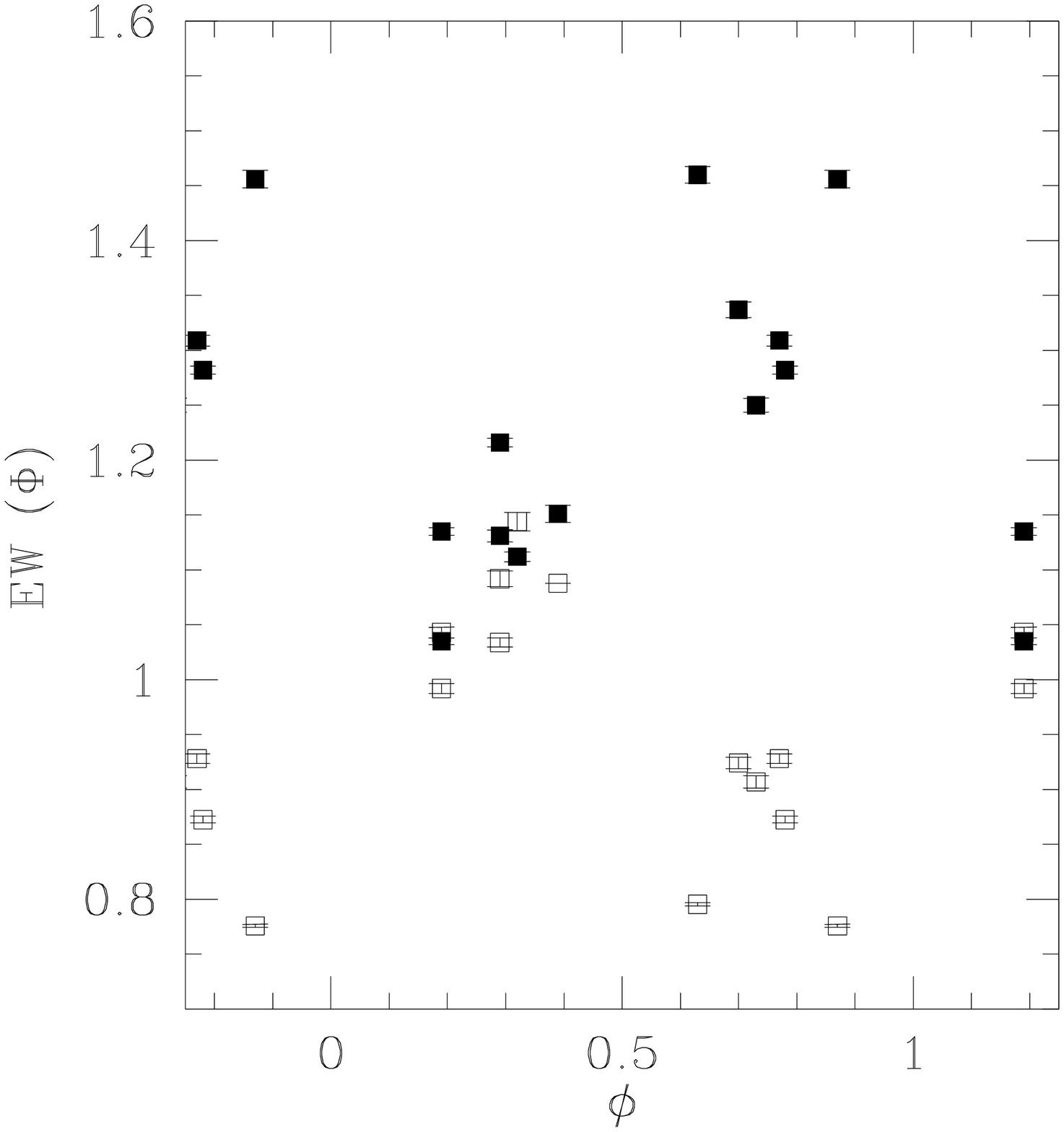}}
  \end{center}
  \caption{Equivalent widths (in \AA) of three helium lines in the spectrum of \object{DH~Cep}, as a function of phase. Filled and open squares represent primary and secondary data, respectively. $\Phi = 0$ corresponds to the phase of conjunction with the secondary star behind. The error bars are formal errors of the Gaussian fits to the line profiles.}
  \label{fig:dhcep_sseffect}
\end{figure*}

\object{DH~Cep} (HD~215835) is a double-lined O~+~O spectroscopic binary that belongs to the young open cluster NGC 7380, in the Perseus arm of our galaxy. \citet{Pearce48} obtained the first orbital solution from optical data. He found a large eccentricity ($e = 0.127 \pm 0.0083$) and an orbital inclination of $i = 62$\degr \, for a period equal to $2.11104 \pm 2 \; 10^{-5}$ days, which leads to very large masses. \citet{HHP76} found small amplitude photometric variations. \citet{LLG86} conducted a more complete photometric campaign on \object{DH~Cep} and confirmed that it is an ellipsoidal variable star. They also added that their light curves imply an eccentricity not larger than 0.05 and an inclination of about 50\degr. A new spectroscopic analysis was performed by \citet{SS94}, using a disentangling method. They found a period very close to the period of \citet{Pearce48}, but with an eccentricity below 0.04. They computed a new light curve from their own results, based on the \citet{LLG86} observations and revised the inclination  downwards to a value of 47\degr. \citet{PGB97} used their tomographic separation method on UV spectra to derive new spectral types for both components: O6~V and O7~V for the primary and the secondary star, respectively.

The early spectral type of \object{DH~Cep}, its short period and the detection of wind material in the system \citep{Corcoran91} make it a good candidate to display the Struve-Sahade effect, even if no detection has been reported yet.

Data were collected between 2002 and 2006 at the Observatoire de Haute-Provence (France) with the Aur\'elie spectrograph mounted on the 1.52m telescope. Aur\'elie was equipped with a 2048\,x\,1024~CCD~EEV~42-20\#3, with a pixel size of 13.5 $\mu$m squared. All spectra were taken with a 600~l/mm grating with a reciprocal dispersion of 16~\AA~mm$^{-1}$, allowing us to achieve a spectral resolution of about 8000 in the blue range. A total of 16 spectra were obtained in the blue part of the visible wavelength domain (from 4450~\AA~to 4900~\AA). The spectra have a mean SNR ratio equal to 430. The data reduction has been done with the MIDAS software and the whole procedure is described in \citet{RDB04}. The journal of the observations is presented in Table~\ref{tab:journal:dhcep}.

\subsection{Orbital solution}

As for \object{HD~165\,052} and \object{HD~159\,176}, we have used the disentangling method (see Fig.~\ref{fig:dhcepdis}) to determine the RVs given in Table~\ref{tab:journal:dhcep}. These values yield the orbital solution of Table~\ref{tab:orbit:dhcep}, for which  the lines considered in the RV calculation are: He~I~$\lambda$~4471, Mg~II~$\lambda$~4481, He~II~$\lambda$~4542, He~II~$\lambda$~4686, He~I~$\lambda$~4713 and H$_{\beta}$. The period was allowed to vary from an initial value of 2.110912 days \citep{SS94}. Fig.~\ref{fig:vrad:dhcep} shows the radial velocity curve as a function of the orbital phase. This solution is close to the solution given by \citet{SS94}, who also used a disentangling method to determine their radial velocities. However, our radial velocities are $\sim$ 15 km~s$^{-1}$ larger. Such a difference could come from the fact that the data of \citet{SS94} did not cover the whole orbital period, or from the different lines considered (they used He~I~$\lambda$~4026, He~II~$\lambda$~4200, He~I~$\lambda$~4471, He~II~$\lambda$~4542 and He~II~$\lambda$~4686).

Using the same approach as for \object{HD~165\,052} and \object{HD~159\,176}, we determined rotational velocities of $162 \pm 10$ and $166 \pm 7$~km~s$^{-1}$ for the primary and secondary component of \object{DH~Cep} respectively. Furthermore, if we suppose that the corresponding angular velocities are equal to the orbital angular velocity of the system, we find an inclination equal to 47\degr, which is exactly the value found by \citet{LLG86} and \citet{HHB96} with photometric data. This indicates a synchronous co-rotation of the stars on a circular orbit ($t_{\mathrm{circ}}$~$<$~21\,000~years, \citet{Tassoul1990}).

\subsection{Spectral type determination}

The measurements of equivalent widths made on the Aurelie spectra of \object{DH~Cep} lead to spectral types O6 and O6.5 for the primary and the secondary star, respectively. However, uncertainties on these numbers (due to the lower resolution of these spectra compared to FEROS spectra, but also to some variations in the He~II~$\lambda$~4542 line, see below) allow the spectral type of the primary to vary from O5.5 to O6.5, and the secondary spectral type from O6 to O6.5. A brightness ratio in our wavelength domain has also been calculated following the method as explained in Sect.~\ref{std165}. We obtain $\frac{L_1}{L_2} = 1.42$ and when we take this number into account, we derive equivalent widths of the He~II~$\lambda$~4686 line equal to 0.66~\AA~for the primary star and 0.82~\AA~for the secondary, which correspond to main sequence stars. When the measurements are performed on the disentangled spectra, the resulting classification is O6~+~O7 system. On the other hand, the comparison of the spectra of Fig. \ref{fig:vrad:dhcep} with the spectra of \citet{WF90} shows that both components should be earlier than an O7 spectral type, with a hotter primary star, thus O6~V~+~O6.5~V.  Both stars have an O((f)) spectral type, as can be seen in Fig.~\ref{fig:dhcepdis} (N~III~$\lambda$$\lambda$~4634, 4640, 4641 lines are in emission).

\subsection{Struve-Sahade effect} 

\begin{table*}
\caption{Intensity variations as a function of phase detected or not (Y or N) for the primary (P) and the secondary (S) of the four systems. The hyphen stands for lines that could not be deblended, because of noise, strong telluric absorption lines, too close components or lines outside the available wavelength domain. (e) means that the line is in emission.}
\label{tab:recap}
\centering
\begin{tabular}{c c c c c c c c c}
\hline
\hline
 &\multicolumn{2}{c}{HD 165\,052}&\multicolumn{2}{c}{HD 100\,213}&\multicolumn{2}{c}{HD 159\,176}&\multicolumn{2}{c}{DH~Cep}\\
Line &P&S&P&S&P&S&P&S\\
\hline
\multicolumn{9}{c}{}\\
He I $\lambda$ 4026&Y&Y&N &N &N &Y & - & -\\
He II $\lambda$ 4200&N &N &-  &- &N &N&- & - \\
He I $\lambda$ 4471&Y&Y&N &N &Y &Y& N & N \\
He II $\lambda$ 4542&N&N&N &N &N &N &Y &Y\\
He II $\lambda$ 4686&N&N&N &N &N &N& Y&Y\\
He I $\lambda$ 4713&N&N&N &N &Y &Y& - & -\\
H$_{\beta}$&- &- &- &- &- &- &Y&Y\\
He I $\lambda$ 4921&N&N&N &N &N &Y& - & - \\
He I $\lambda$ 5016&N&N&N &N &N &N & - & -\\
He II $\lambda$ 5412&N&N&N &N &N &N & - & -\\
C III $\lambda$ 5696&N (e)&N (e)&- &- &N (e) & N (e) & - & -\\
C IV $\lambda$$\lambda$ 5801,5812&N&N&- &- &N & N & - & -\\
He I $\lambda$ 5876&N&N &- &Y &Y &Y & - & -\\
\hline
\end{tabular}
\end{table*}

Because of the small wavelength domain of the Aur\'elie spectra, only a few lines could be checked for the Struve-Sahade effect: He~I~$\lambda$~4471, He~II~$\lambda$$\lambda$~4542,~4686 and H$_{\beta}$. Among these lines, only the first one does not show any variations. Indeed, as can be seen in Fig. \ref{fig:dhcep_sseffect}, the equivalent widths of the He~II lines and of H$_{\beta}$ vary with the orbital phase but in different ways. The equivalent width of the He~II~$\lambda$~4542 line seems to decrease when the star is receding, with a minimum during the conjunction when the star is behind its companion, and to increase during the other half cycle (Fig. \ref{subfig:dhcep4542}). This behavior is similar to the He~I~$\lambda$$\lambda$~4026, 4471 lines of \object{HD~100\,213}. Since \object{DH~Cep} is also an ellipsoidal variable \citep{LLG86} this behavior indicates that He~II~$\lambda$~4542 is preferentially formed at the back side of the stars. For the two other lines, the details of the variation are less clear. However, we can say that globally, the equivalent width of the H$_{\beta}$ line is larger by 20\% when the star is approaching, for both components. The same holds for the equivalent width of the  primary component of the He~II~$\lambda$~4686 line when the primary star is approaching. For the secondary component of He~II~$\lambda$~4686, the variation is only 10\%. In order to estimate errors due to noise or poor normalisation, we measured the standard deviation of the equivalent width of a DIB ($\lambda = 4762$~\AA) and found $\sigma = 2$\%. Finally, no DIB exists in the vicinity of the measured lines that could bias the measurements.

\section{Discussion and conclusions}

We have presented a spectroscopic analysis of four very different early-type binary systems, based on observations obtained with the ESO echelle spectrograph FEROS and the OHP Aur\'elie instrument. We first used the data to determine new orbital solutions and spectral types. It was necessary to know as well as possible the fundamental parameters of the systems in order to study the Struve-Sahade effect.  The conclusion of \citet{BGR99} that the effect arises from different mechanisms and that each star has its own 'tale' seems to apply also to \object{HD~165\,052}, \object{HD~100\,213}, \object{HD~159\,176} and \object{DH~Cep}. Table~\ref{tab:recap} shows a summary of the results obtained for each system. It indicates which lines present variations during the orbital cycle and if they appear in both stars.

In the first of these systems, \object{HD~165\,052}, some variations were detected but only in two lines, He~I~$\lambda$~4026 and He~I~$\lambda$~4471. Indeed, the ratio between the primary and secondary equivalent width is lower than 1 when the secondary star is approaching, and greater than 1 half an orbital cycle later. This can be compared to the behavior of \object{HD~93\,403}, an O5.5~I + O7~V system that is also known to display the Struve-Sahade effect \citep{RSG00,RVS02}. In this latter system the ratio $EW_P$/$EW_S$ of He~I~$\lambda$~4471, C~IV~$\lambda$$\lambda$~5801, 5812 and C~III~$\lambda$~5696 is less than one when the secondary star is approaching, and greater or equal to one when it is receding. The difference between the two systems is that in the case of \object{HD~93\,403}, only the secondary lines presented some variations, while in \object{HD~165\,052} both stars are affected.

In the case of \object{HD~100\,213}, there does not seem to be a genuine Struve-Sahade effect. The variations observed in the He~I lines are rather linked to the fact that the lines are formed over a limited part of the stellar surface that is not visible during the entire orbital cycle. In fact, the behavior of these lines, both in equivalent width and radial velocity, suggests that they are preferentially formed on the rear side of the stars, away from the radiations of the companion. Indeed, in this contact binary system, it seems likely that mutual heating is responsible for the non-uniform He~I strength over the stellar surface.

In \object{HD~159\,176}, almost all He~I lines seem to vary. For two of them (He~I~$\lambda$$\lambda$~4026, 4921) only the secondary lines are modified during the orbital cycle, while for the other ones (He~I~$\lambda$$\lambda$~4471, 4713, 5876) both primary and secondary lines are affected. The ratio $EW_P$/$EW_S$ increases just before the secondary eclipse, and decreases before the primary eclipse, as can be seen in the O7~III(f)+O7~III(f) system \object{HD~152\,248} \citep{SRG01} in the He~I~$\lambda$$\lambda$~4713, 4921 lines. The situation is again similar to \object{HD~100\,213} and mutual heating would once more lead to an inhomogeneous distribution of the He~I line intensity over the surface of the star. 

\object{DH~Cep} also shows equivalent width variations in three of the four examined spectral lines. This massive binary is an ellipsoidal variable system \citep{LLG86}, and as for \object{HD~100\,213}, the variations in He~II~$\lambda$~4542 are probably due to the fact that the line is formed preferentially over the rear side of the star. While it seems easy to explain a somewhat higher temperature and hence a weaker He~I line on the front side of the stars by heating effects, explaining an enhanced He~II line on the rear of the stars in \object{DH~Cep} is harder. The behavior of  He~II~$\lambda$~4686 and H$_{\beta}$ is different. Indeed, the equivalent widths of these two lines are larger when the star is approaching, in agreement with the definition of the Struve-Sahade effect, except that it is not only the secondary star that is concerned by these variations.

In summary, the behavior of these four systems differs from what is expected for the 'standard' Struve-Sahade effect. First, only several lines vary and they are not the same for each system. The line variations of at least two of the studied binaries (\object{HD~100\,213} and \object{HD~159\,176}) can be explained by the fact that the line formation is not homogeneous over the whole surface of the star. \object{HD~165\,052} follows more or less the definition of the S-S effect, but it is now clear that both spectra are affected and not the secondary star spectrum only. Line variations of \object{DH~Cep}, in which no S-S had ever been detected, have both behaviors depending on the considered line.

The binaries in the present sample are composed of main sequence stars, and a similar study will be presented in a subsequent paper (paper II) for giant and supergiant objects (\object{HD~47\,129}, \object{29~CMa}, \object{AO~Cas}). A comparison between these observed data and synthetic spectra is needed to understand the cause of these variations during the orbital cycle. This will be the subject of a forthcoming paper (paper III).

\begin{acknowledgements}
N. L. acknowledges the financial support of the Belgian "Fonds pour la Formation \`a la Recherche dans l'Industrie et dans l'Agriculture". This research is supported in part through the XMM/INTEGRAL PRODEX contract. 
\end{acknowledgements}

\bibliographystyle{aa}

\bibliography{/home/linder/Bib_hugues/natacha.bib}

\begin{thebibliography}{49}
\expandafter\ifx\csname natexlab\endcsname\relax\def\natexlab#1{#1}\fi

\bibitem[{{Andersen} \& {Gronbech}(1975)}]{AG75}
{Andersen}, J. \& {Gronbech}, B. 1975, \aap, 45, 107

\bibitem[{{Arias} {et~al.}(2002){Arias}, {Morrell}, {Barb{\'a}}, {Bosch},
  {Grosso}, \& {Corcoran}}]{AMB02}
{Arias}, J.~I., {Morrell}, N.~I., {Barb{\'a}}, R.~H., {et~al.} 2002, \mnras,
  333, 202

\bibitem[{{Bagnuolo} {et~al.}(1999){Bagnuolo}, {Gies}, {Riddle}, \&
  {Penny}}]{BGR99}
{Bagnuolo}, W.~G., {Gies}, D.~R., {Riddle}, R., \& {Penny}, L.~R. 1999, \apj,
  527, 353

\bibitem[{{Conti}(1973)}]{Conti73}
{Conti}, P.~S. 1973, \apj, 179, 161

\bibitem[{{Conti}(1974)}]{Conti74}
{Conti}, P.~S. 1974, \apj, 187, 539

\bibitem[{{Conti} \& {Alschuler}(1971)}]{CA71}
{Conti}, P.~S. \& {Alschuler}, W.~R. 1971, \apj, 170, 325

\bibitem[{{Conti} {et~al.}(1975){Conti}, {Cowley}, \& {Johnson}}]{CCJ75}
{Conti}, P.~S., {Cowley}, A.~P., \& {Johnson}, G.~B. 1975, \pasp, 87, 327

\bibitem[{{Corcoran}(1991)}]{Corcoran91}
{Corcoran}, M.~F. 1991, \apj, 366, 308

\bibitem[{{Corcoran}(1996)}]{Corcoran96}
{Corcoran}, M.~F. 1996, in Revista Mexicana de Astronomia y Astrofisica
  Conference Series, ed. V.~{Niemela} \& N.~{Morrell}, 54

\bibitem[{{De Becker} {et~al.}(2004){De Becker}, {Rauw}, {Pittard}, {Antokhin},
  {Stevens}, {Gosset}, \& {Owocki}}]{DBRP04}
{De Becker}, M., {Rauw}, G., {Pittard}, J.~M., {et~al.} 2004, \aap, 416, 221

\bibitem[{{Gayley}(2002)}]{Gayley02}
{Gayley}, K.~G. 2002, in ASP Conf. Ser. 260: Interacting Winds from Massive
  Stars, ed. A.~F.~J. {Moffat} \& N.~{St-Louis}, 583

\bibitem[{{Gayley} {et~al.}(2007){Gayley}, {Townsend}, {Parsons}, \&
  {Owocki}}]{GTP07}
{Gayley}, K.~G., {Townsend}, R., {Parsons}, J., \& {Owocki}, S. 2007, in ASP
  Conference Series, in press

\bibitem[{{Gies} {et~al.}(1997){Gies}, {Bagnuolo}, \& {Penny}}]{GBP97}
{Gies}, D.~R., {Bagnuolo}, W.~G., \& {Penny}, L.~R. 1997, \apj, 479, 408

\bibitem[{{Gonz{\'a}lez} \& {Levato}(2006)}]{GL06}
{Gonz{\'a}lez}, J.~F. \& {Levato}, H. 2006, \aap, 448, 283

\bibitem[{{Gray}(2005)}]{Gray2005}
{Gray}, D.~F. 2005, {The Observation and Analysis of Stellar Photospheres}
  (Cambridge University Press)

\bibitem[{{Hilditch} {et~al.}(1996){Hilditch}, {Harries}, \& {Bell}}]{HHB96}
{Hilditch}, R.~W., {Harries}, T.~J., \& {Bell}, S.~A. 1996, \aap, 314, 165

\bibitem[{{Hill} {et~al.}(1976){Hill}, {Hilditch}, \& {Pfannenschmidt}}]{HHP76}
{Hill}, G., {Hilditch}, R.~W., \& {Pfannenschmidt}, F.~L. 1976, Publications of
  the Dominion Astrophysical Observatory Victoria, 15, 1

\bibitem[{{Howarth} {et~al.}(1997){Howarth}, {Siebert}, {Hussain}, \&
  {Prinja}}]{HSH97}
{Howarth}, I.~D., {Siebert}, K.~W., {Hussain}, G.~A.~J., \& {Prinja}, R.~K.
  1997, \mnras, 284, 265

\bibitem[{{Lines} {et~al.}(1986){Lines}, {Lines}, {Guinan}, \&
  {Robinson}}]{LLG86}
{Lines}, H.~C., {Lines}, R.~D., {Guinan}, E.~F., \& {Robinson}, C.~R. 1986,
  Information Bulletin on Variable Stars, 2932, 1

\bibitem[{{Martins} {et~al.}(2005){Martins}, {Schaerer}, \& {Hillier}}]{MSH05}
{Martins}, F., {Schaerer}, D., \& {Hillier}, D.~J. 2005, \aap, 436, 1049

\bibitem[{{Mathys}(1988)}]{Mathys88}
{Mathys}, G. 1988, \aaps, 76, 427

\bibitem[{{Mathys}(1989)}]{Mathys89}
{Mathys}, G. 1989, \aaps, 81, 237

\bibitem[{{Morrison}(1975)}]{Morrison75}
{Morrison}, N.~D. 1975, \apj, 200, 113

\bibitem[{{Morrison} \& {Conti}(1978)}]{MC78}
{Morrison}, N.~D. \& {Conti}, P.~S. 1978, \apj, 224, 558

\bibitem[{{Pachoulakis}(1996)}]{Pachoulakis96}
{Pachoulakis}, I. 1996, \mnras, 280, 153

\bibitem[{{Pearce}(1948)}]{Pearce48}
{Pearce}, J.~A. 1948, \aj, 54, 135

\bibitem[{{Penny} {et~al.}(1997){Penny}, {Gies}, \& {Bagnuolo}}]{PGB97}
{Penny}, L.~R., {Gies}, D.~R., \& {Bagnuolo}, Jr., W.~G. 1997, \apj, 483, 439

\bibitem[{{Rauw} \& {De Becker}(2004)}]{RDB04}
{Rauw}, G. \& {De Becker}, M. 2004, \aap, 421, 693

\bibitem[{{Rauw} {et~al.}(2000){Rauw}, {Sana}, {Gosset}, {Vreux}, {Jehin}, \&
  {Parmentier}}]{RSG00}
{Rauw}, G., {Sana}, H., {Gosset}, E., {et~al.} 2000, \aap, 360, 1003

\bibitem[{{Rauw} {et~al.}(2002){Rauw}, {Vreux}, {Stevens}, {Gosset}, {Sana},
  {Jamar}, \& {Mason}}]{RVS02}
{Rauw}, G., {Vreux}, J.-M., {Stevens}, I.~R., {et~al.} 2002, \aap, 388, 552

\bibitem[{{Sahade}(1959)}]{Sahade59}
{Sahade}, J. 1959, \pasp, 71, 151

\bibitem[{{Sana} {et~al.}(2006){Sana}, {Gosset}, \& {Rauw}}]{SGR06b}
{Sana}, H., {Gosset}, E., \& {Rauw}, G. 2006, \mnras, 371, 67

\bibitem[{{Sana} {et~al.}(2003){Sana}, {Hensberge}, {Rauw}, \&
  {Gosset}}]{SHR03}
{Sana}, H., {Hensberge}, H., {Rauw}, G., \& {Gosset}, E. 2003, \aap, 405, 1063

\bibitem[{{Sana} {et~al.}(2001){Sana}, {Rauw}, \& {Gosset}}]{SRG01}
{Sana}, H., {Rauw}, G., \& {Gosset}, E. 2001, \aap, 370, 121

\bibitem[{{Sim{\'o}n-D{\'{\i}}az} \& {Herrero}(2007)}]{SDH07}
{Sim{\'o}n-D{\'{\i}}az}, S. \& {Herrero}, A. 2007, ArXiv Astrophysics e-prints

\bibitem[{{Stickland}(1997)}]{Stickland97}
{Stickland}, D.~J. 1997, The Observatory, 117, 37

\bibitem[{{Stickland} {et~al.}(1993){Stickland}, {Koch}, {Pachoulakis}, \&
  {Pfeiffer}}]{SKP93}
{Stickland}, D.~J., {Koch}, R.~H., {Pachoulakis}, I., \& {Pfeiffer}, R.~J.
  1993, The Observatory, 113, 204

\bibitem[{{Stickland} {et~al.}(1997){Stickland}, {Lloyd}, \& {Koch}}]{SLK97}
{Stickland}, D.~J., {Lloyd}, C., \& {Koch}, R.~H. 1997, The Observatory, 117,
  295

\bibitem[{{Stickland} {et~al.}(1995){Stickland}, {Lloyd}, {Koch}, \&
  {Pachoulakis}}]{SLK95}
{Stickland}, D.~J., {Lloyd}, C., {Koch}, R.~H., \& {Pachoulakis}, I. 1995, The
  Observatory, 115, 317

\bibitem[{{Struve}(1937)}]{Struve37}
{Struve}, O. 1937, \apj, 85, 41

\bibitem[{{Struve}(1950)}]{Struve50}
{Struve}, O. 1950, {Stellar evolution, an exploration from the observatory.}
  (Princeton University Press)

\bibitem[{{Sturm} \& {Simon}(1994)}]{SS94}
{Sturm}, E. \& {Simon}, K.~P. 1994, \aap, 282, 93

\bibitem[{{Tassoul}(1990)}]{Tassoul1990}
{Tassoul}, J.-L. 1990, \apj, 358, 196

\bibitem[{{Terrell} {et~al.}(2003){Terrell}, {Munari}, {Zwitter}, \&
  {Nelson}}]{TMZ03}
{Terrell}, D., {Munari}, U., {Zwitter}, T., \& {Nelson}, R.~H. 2003, \aj, 126,
  2988

\bibitem[{{Thomas}(1975)}]{Thomas75}
{Thomas}, J.~C. 1975, in Bulletin of the American Astronomical Society, 533

\bibitem[{{Trumpler}(1930)}]{Trumpler30}
{Trumpler}, R.~J. 1930, \pasp, 42, 342

\bibitem[{{van Altena} \& {Jones}(1972)}]{VAJ72}
{van Altena}, W.~F. \& {Jones}, B.~F. 1972, \aap, 20, 425

\bibitem[{{Walborn} \& {Fitzpatrick}(1990)}]{WF90}
{Walborn}, N.~R. \& {Fitzpatrick}, E.~L. 1990, \pasp, 102, 379

\bibitem[{{Wolfe} {et~al.}(1967){Wolfe}, {Horak}, \& {Storer}}]{WHS67}
{Wolfe}, Jr., R.~H., {Horak}, H.~G., \& {Storer}, N.~W. 1967 (Modern
  astrophysics.~A memorial to Otto Struve), 251

\end{thebibliography}

\end{document}